\documentclass[prd,twocolumn,nofootinbib,superscriptaddress,letterpaper,floatfix]{revtex4-2}

\makeatletter
\let\@bibdataout@init\@empty
\def\@bibdataout@rev{%
  \if@filesw
    \immediate\write\@auxout{\string\citation{REVTEX42Control,apsrev42Control}}%
  \fi
}
\makeatother

\usepackage{amsmath,amssymb,bm,dcolumn}
\usepackage{xspace}
\usepackage{url}
\usepackage{graphicx}
\usepackage{multirow}
\usepackage{tikz}
\usetikzlibrary{shapes.geometric,arrows}

\usepackage{xcolor}
\usepackage{listings}
\usepackage{orcidlink}

\usepackage{hyperref}
\hypersetup{colorlinks=true,allcolors=blue}
\setcitestyle{numbers,sort&compress,comma}
\definecolor{amber}{rgb}{1.0,0.75,0.0}
\definecolor{amaranth}{rgb}{0.9,0.17,0.31}
\definecolor{orangeCustom}{rgb}{1.0,0.5,0.0}

\tikzset{
  startstop/.style = {circle,rounded corners,minimum
  width=1cm,minimum height=1cm,text centered,draw=black,fill=red!30},
  io/.style        = {trapezium,trapezium left angle=70,trapezium
    right angle=110,minimum width=2cm,minimum height=1cm,text
  centered,draw=black,fill=blue!30},
  process/.style   = {rectangle,minimum width=2cm,minimum
  height=1cm,text centered,draw=black,fill=orangeCustom!30},
  decision/.style  = {diamond,minimum width=2cm,minimum
  height=1cm,text centered,draw=black,fill=green!30},
  arrow/.style     = {thick,->,>=stealth}
}

\newcommand{\RIT}{Center for Computational Relativity and
Gravitation, Rochester Institute of Technology, Rochester, New York 14623, USA}
\newcommand{\CITA}{Canadian Institute for Theoretical Astrophysics,
60 St George St, University of Toronto, Toronto, ON M5S 3H8, Canada}

\newcommand{\llmfig}{Generated with LLM assistance per Appendix~\ref{app:llm}.}
\newcommand{\ForInternalReference}[1]{}
\newcommand{\SkipForEarlyCirculation}[1]{}

\newcommand{\SkipPP}[1]{}
\newcommand{\optional}[1]{}

\newcommand{\gwk}{\textsc{GWKokab}\xspace}

\newcommand{\rate}{\mathcal{R}}
\newcommand{\pvec}{\Lambda} 
\newcommand{\svec}{\lambda} 

\newcommand{\comp}{\boldsymbol{\rho}}

\newcommand{\prob}{\operatorname{P}}
\newcommand{\mean}{\boldsymbol{\mu}}
\newcommand{\cov}{\boldsymbol{\Sigma}}

\begin{document}
\renewcommand{\arraystretch}{1.2}

\title{Narrow Population Inference Enhanced by Analytical Likelihood Models}

\author{M. Qazalbash \orcidlink{0009-0002-9640-1185}}
\affiliation{\RIT}

\author{S. Mehra}
\affiliation{\RIT}

\author{M. Zeeshan \orcidlink{0000-0002-6494-7303}}
\email{m.zeeshan5885@gmail.com}
\affiliation{\RIT}

\author{V. Delfavero \orcidlink{0000-0001-7099-765X}}
\affiliation{\CITA}

\author{R. O'Shaughnessy \orcidlink{0000-0001-5832-8517}}
\affiliation{\RIT}


\begin{abstract}
	The growing catalog of gravitational-wave events has revealed
	substantial diversity in the properties of compact-binary mergers.
	However, commonly used population-inference methods based on
	discrete posterior samples can struggle to constrain narrow
	population features, resulting in biased or unstable estimates of
	population hyperparameters. We first demonstrate this limitation
	using a toy population model by comparing parameter recovery with a
	continuous likelihood model against discrete approximations
	constructed from $10^3$, $10^4$, and $10^5$ samples. We then perform
	the same comparison using synthetic eccentric and multisource
	populations introduced in previous studies. Although the continuous
	and discrete approaches yield broadly consistent results, the
	continuous approximation more accurately recovers the parameters of
	narrow simulated populations. In particular, while both methods
	produce similar mass distributions, appreciable differences arise
	for narrowly distributed parameters such as spin and eccentricity.
	Our results indicate that the continuous approach provides more
	reliable inference for spin and eccentricity, whose narrow
	population distributions can be inadequately represented by finite
	sample sets. In conditional comparisons using GWTC-4 and GWTC-5,
	both event representations recover the primary-mass peak near
	$10\,M_\odot$, and a variance cut driven by the selection injections
	preferentially excludes narrower peaks. Continuous likelihood
	models therefore offer a valuable
	tool for improving population inference and extracting more robust
	information about the formation and evolution of compact-binary
	systems.

\end{abstract}

\maketitle

\section{Introduction}\label{sec:intro}

Since the first direct detection of gravitational waves (GWs) by the
Laser Interferometer Gravitational-Wave Observatory (LIGO)
\cite{2016PhRvL.116f1102A}, the catalog of GW sources
\cite{2019PhRvX...9c1040A,2021PhRvX..11b1053A,2024PhRvD.109b2001A,2023PhRvX..13d1039A,2025arXiv250818082T}
observed by the LVK detectors
\cite{2015CQGra..32g4001L,2015CQGra..32b4001A,2021PTEP.2021eA101A} has
been steadily growing, a trend expected to continue as detector
sensitivities improve
\cite{2015CQGra..32g4001L,2019NatAs...3...35K,2021PTEP.2021eA101A,2025arXiv250818081T,2020LRR....23....3A}.
These observations have enabled many important investigations,
including studies of compact-object formation
\cite{Vitale_2017,Abbott_2016,Zevin_2017,MANDEL20221,2009LRR....12....2S,
	Thorne1977,2016Natur.534..512B,2022ApJ...940..171R,2020ApJ...903L...5R,2021ApJ...921L..43Z},
constraints on the structure and evolution of the Universe
\cite{2013arXiv1307.2638N,Holz_2005,PhysRevLett.108.091101,2017Natur.551...85A,Christensen_2019,PhysRevLett.129.061102,Gair_2023},
tests of general relativity in the strong-gravity regime
\cite{PhysRevD.85.024041,2014LRR....17....4W,2016phrvl.116v1101a,PhysRevLett.123.011102,PhysRevD.112.084080,2023PhRvD.108b4043M,2023PhRvD.107d4020M,2021PhRvD.103l2002A,2019PhRvL.123l1101I},
and probes of the equation of state of matter at supranuclear densities
\cite{Baym_2018,2020NatAs...4..625C,2020Sci...370.1450D,PhysRevD.101.123007}.

In recent years, population analyses of compact binaries have
increasingly been used to characterize the distributions of their
masses, spins, tilt angles, eccentricities, and distances
\cite{2022LRR....25....1M,2022phrvd.106j3019t,2017natur.548..426f,2018apj...868..140t,2019PhRvD.100d3012W,2025arXiv250818083T,Vitale2020}.
For instance, the sharp decline observed in the binary black hole
(BBH) mass spectrum above 50 $M_\odot$
\cite{Fishbach_2017,Edelman_2021,physrevx.13.011048} points to the
pair-instability mechanism as a key factor limiting the masses of
stellar cores
\cite{1964ApJS....9..201F,1967PhRvL..18..379B,2002ApJ...567..532H,2003ApJ...591..288H,2015ASSL..412..199W,2016A&A...594A..97B,2017ApJ...836..244W,2019ApJ...882...36M,2020A&A...640A..56R}.
The mass scale of this cutoff offers valuable insights into nuclear
reaction rates in massive stars \cite{Farmer_2020}. Similarly, the
observed peak in the BBH mass spectrum near $30$-$40\,M_\odot$
\cite{2021apj...913l...7a,physrevx.13.011048,2023ApJS..267...29A} is
attributed to the accumulation of systems formed through pulsational
pair-instability supernovae
\cite{2017ApJ...836..244W,2018ApJ...856..173T}, with the peak's
location largely independent of stellar metallicity and redshift
\cite{2019ApJ...887...53F}. Additional substructures in the BBH mass
spectrum, such as peaks and troughs superimposed on a smooth power
law
\cite{2021ApJ...913L..19T,2022ApJ...924..101E,2023ApJ...946...16E,2022ApJ...928..155T,2026arXiv260600234P},
have helped constrain the relative contributions of various formation
channels to the BBH population. Population-level analyses
of BBH redshifts using the Gravitational-Wave Transient Catalog
(GWTC) reveal that the BBH merger
rate increases with redshift
\cite{2018apj...863l..41f,2021ApJ...912...98F,2023MNRAS.523.4539K,2022ApJ...931...17V,2022ApJ...940..184V,2026arXiv260602318P},
providing insights into the metallicity evolution and star formation
history of the Universe.

Spin distributions also provide critical information about the
formation and evolution of compact binaries. Numerous studies suggest
that most black holes are born with very low natal spins due to
efficient angular momentum transport within stars
\cite{2019ApJ...881L...1F,2019MNRAS.485.3661F,2019MNRAS.488.4338M}, a
trend supported by observations in gravitational wave catalogs
\cite{2025arXiv250818083T,2021PhRvD.104h3010R,2023Univ....9..507P,2023PhRvD.108j3009G,2022ApJ...937L..13C,adamcewicz2025both}.
In contrast, black holes in X-ray binary systems often exhibit high
spins \cite{2022MNRAS.511.3951F,2019ApJ...870L..18Q,MILLER20151}, and
some binary components appear to have significant spins. These
observations appear to conflict with the low spins inferred from GW
events.
Several studies offer potential explanations for this apparent
discrepancy \cite{2022ApJ...929L..26F,2024A&A...690A..21B,2021ApJ...921L..15G},
suggesting that differences in formation channels may underlie these
observations. However, even with existing samples, this remains a
subject of active debate in the literature. Therefore, improving
population inference of compact binaries is crucial for advancing our
understanding of many astrophysical processes, particularly subtle
and narrow features in their distributions.

Hierarchical Bayesian inference
\cite{ThraneTalbot2019,2019MNRAS.486.1086M,Vitale2020} lies at the
core of population analyses. It provides a Bayesian framework to
constrain the rate densities that define a parametric population model
while accounting for selection effects and measurement uncertainties.
Bayesian inference applied to GW data yields posterior samples for
each event. These parameter-estimation (PE) samples can be reweighted
according to a chosen population model. Treating multiple detections
as independent realizations of an inhomogeneous Poisson process
allows them to be combined hierarchically to construct the likelihood
of population-level parameters given the entire data set
\cite{2004AIPC..735..195L,2019MNRAS.486.1086M,2019PhRvD.100d3012W}.
The inference proceeds via importance sampling
\cite{910e8e7d-2e70-317c-adf6-e21cd50755a1} to estimate the target
distribution proportional to the desired population-level posterior.
However, since importance sampling relies on a finite set of samples,
it only approximates the true target distribution.

Previous studies
\cite{2022arxiv220400461e,2025arXiv250907221H,2023MNRAS.526.3495T,2025PhRvD.111l3049M,2022ApJ...927...76T}
have demonstrated that inference can be performed using a fixed,
large number of posterior samples, independent of catalog size,
provided the number of injections scales linearly with the catalog
size. Yet, when target distributions are marginalized over
approximate Monte Carlo uncertainties, additional samples are needed
to avoid bias. While this scaling is manageable for small catalogs,
it becomes computationally challenging for larger datasets,
potentially leading to biased inference with inflated uncertainties
for narrow populations.

Following previous work
\cite{2020arxiv200101747w,2015PhRvD..91d3002L,2018PhRvD..98f3004C,2020ApJ...888...12M},
we address these challenges by modeling single-event likelihoods
with multivariate Gaussian densities that continuously approximate
discrete posterior samples. We investigate the effects of these
approximations on several synthetic catalogs. The method is
implemented in the population-inference engine \gwk
\cite{krnm-3vrf,git_gwkokab}. It models each single-event likelihood
as a multivariate Gaussian and uses a carefully chosen proposal
distribution for importance sampling. Tests on broad and narrow
synthetic catalogs show that the method produces improved estimates
of population hyperparameters with reduced variance compared with
the standard discrete-sample approach.

This paper is organized as follows. Section \ref{sec:methods} reviews
hierarchical Bayesian inference and describes the implementation of
continuous likelihoods in \gwk, the generation of synthetic
populations and discrete PE samples, and their continuous
approximations. Section \ref{sec:analyses} compares discrete and
continuous likelihood representations using synthetic data and tests
recovery of the GWTC-4.0 low-mass peak with real events. Finally,
Sec. \ref{sec:conclude}
summarizes key findings and outlines directions for future work.
\section{Methods}\label{sec:methods}
\subsection{Review of Hierarchical Bayesian Inference (HBI)}

To infer compact-binary populations, we adopt the Bayesian Parametric
Model (BPM) formalism introduced in previous work and implemented in
the population-inference engine \gwk. Within this framework, binaries
with intrinsic parameters $\svec$ merge at a rate
$\comp(\svec\mid\pvec)$ per unit volume, source-frame time, and
source-parameter volume. This rate depends on the population
hyperparameters $\pvec$. The BPM inputs are (a) the underlying
parameterized population model $\comp(\pvec)$ and
(b) the expected number of detections, $\mu(\pvec)$, for a given
experiment.
In terms of these ingredients, we express the likelihood of an
astrophysical BBH population with parameters $\pvec$
as a conventional inhomogeneous Poisson process:
\begin{equation}\label{eq:likelihood}
	\mathcal{L}( \pvec) \propto
	e^{-\mu(\pvec)}
	\prod_{j=1}^N
	\int \ell_j(\svec) \comp(\svec|\pvec)  \sqrt{g_\svec} \mathrm{d}\svec,
\end{equation}
Here, $\sqrt{g_\svec}$ characterizes the four-volume and parameter-space
measure relative to which $\comp$ is defined, as described in our
previous work \cite{krnm-3vrf}; $\mu(\pvec)$ is the expected number of
detections for population parameters $\pvec$; and
$\ell_j(\svec)=p(d_j\mid\svec)$ is the likelihood of the data $d_j$
for the $j$th detection, given binary parameters $\svec$.

Evaluating this likelihood requires two types of integrals.
The first determines the single-event marginal likelihoods,
$\int \ell_j(\svec)\comp(\svec\mid\pvec)\sqrt{g_\svec}\,
\mathrm{d}\svec$. The second determines the expected number of
events, which we write as
\begin{align}
	\label{eq:mu}
	\mu(\pvec)= \int P_{\mathrm{det}} (\svec;z)\cdot
	\comp(\svec\mid\pvec)\sqrt{ g_{\svec}}
	d \svec.
\end{align}
Here, $P_{\rm det}$ is the detection probability for a source with
intrinsic parameters $\svec$ at redshift $z$.
Following established practice, when appropriate, we evaluate this
integral for populations whose event rate is uniform per unit
comoving volume and source-frame time in terms of a sensitive
four-volume $\mathrm{VT}(\lambda)$, as reviewed in
Appendix \ref{ap:VT}. Our notation and approach are described further in
\cite{krnm-3vrf}.

In principle, both types of integrals can be computed by quadrature
for all events simultaneously, provided that the inputs
$\ell_j(\svec)$ and $P_{\rm det}$ are known exactly. In practice, each
integral can be evaluated in two ways, corresponding to different
strategies for representing the imperfect performance of GW searches
and parameter-estimation analyses. The integral defining
$\mu(\pvec)$ can be evaluated using a set of injections
\cite{2025PhRvD.112j2001E}, represented by a large but fixed set of
evaluation points $\svec_k$ and draw probabilities
$\prob_{\rm draw}(\svec_k)$.
Alternatively, continuous approximations can be derived that encode
search performance, as outlined in previous work
\cite{krnm-3vrf,pop-models-aps-2021-vt,2024PhRvD.110l3041C,2024CQGra..41l5002G,2023PhRvD.108d3011E}
and Appendix \ref{ap:VT}.

Similarly, the marginal evidence for each event can be calculated
from a fiducial analysis using either a discrete or a continuous
approximation:

\begin{equation}
	\label{eq:marginal_likelihood}
	{\cal Z}_j(\pvec) \equiv \int \sqrt{g_\svec}  d\svec\;
	\ell_j(\svec) \comp(\svec|\pvec).
\end{equation}

Discrete methods evaluate the integral using a sampling distribution
$\pi_s$ and samples $\svec_\alpha$:
\begin{equation}
	\label{eq:marginal_likelihood_approximation}
	{\cal Z}_j(\pvec) \approx \frac{1}{N} \sum_\alpha
	\frac{\ell_j(\svec_\alpha) \comp(\svec_\alpha|\pvec) \sqrt{g_\svec}
	}{\pi_s(\svec_\alpha)}
\end{equation}
The most widely used discrete method for hierarchical inference
employs a particularly efficient technique in which
samples are drawn from the posterior distribution relative to some
reference prior $\pi_{\rm ref}$, obviating the need to access the
original likelihood.
\begin{equation}
	\label{eq:marginal_likelihood_discrete_standard}
	{\cal Z}_j(\pvec) \approx \frac{Z}{N} \sum_\alpha \frac{
		\comp(\svec_\alpha|\pvec) \sqrt{g_\svec} }{\pi_{\rm ref}(\svec_\alpha)}
\end{equation}
Alternatively, if a continuous model for $\ell_j(\svec)$ is
available, the integrals ${\cal Z}_j$ can be evaluated using standard
numerical quadrature techniques, including adaptive Monte Carlo
integration
\cite{2022arXiv220514154D,2022APS..APRX16004D,2021arXiv210713082D,
	2020arxiv200101747w}.

Nearly all publications characterizing GW events provide posterior
samples and describe or provide the corresponding reference prior
distribution. Given continuous representations of
$P_{\rm det}(\svec)$ and $\ell_j(\svec)$, however, these integrals can
instead be evaluated using other quadrature methods
\cite{2020arxiv200101747w}. Some parameter-inference engines, such as
RIFT, already produce and export an estimate of the marginal
likelihood as a data product using either Gaussian-process or random-
forest interpolation. For high-amplitude signals, the RIFT marginal
likelihood can often be approximated by a Gaussian in suitable
coordinates.
For conventional MCMC engines that report only posterior samples, the
likelihood can sometimes be represented by a carefully tuned density
approximation, such as a Gaussian kernel density estimate. Finally,
for simple investigations that do not require end-to-end parameter
inference, a suitable approximate marginal likelihood
$\ell_j(\svec)$ can be generated using a Fisher-matrix approximation,
as implemented in
the \textsc{synthetic-PE-posteriors} library.

\subsection{Optimized Monte Carlo Integration for Narrow Likelihoods}

Following
\cite{2022arXiv220514154D,2022APS..APRX16004D,2021arXiv210713082D},
we approximate the marginal likelihood as a multivariate Gaussian
distribution with nonzero support over the hypercube
$[\mathbf{x}_0,\mathbf{x}_1]$, mean vector $\mean$, and covariance
matrix $\cov$. The approximation may be constructed in a different
set of parameters, $\svec'$, obtained by mapping $\svec$ through
$\Psi$. Under these assumptions, Eq. \ref{eq:marginal_likelihood}
becomes

\begin{equation}
	{\cal Z}(\pvec) = \int
	\mathcal{N}\left(\Psi(\svec)|\mean,\cov\right)
	\comp(\svec|\pvec) \sqrt{g_\svec}  d\svec
\end{equation}
We then change variables and use Monte Carlo integration over the
hypercube to approximate the integral:
\begin{equation}\label{eq:marginal_likelihood_analytical}
	{\cal Z}(\pvec) = \int_{[\mathbf{x}_0,\mathbf{x}_1]}
	\mathcal{N}\left(\Psi|\mean,\cov\right)
	\comp\left( \lambda(\Psi) | \pvec \right) \sqrt{g_\svec}
	\left|
	\frac{\partial \svec }{\partial\Psi}
	\right|
	d\Psi
\end{equation}
where $\left|\partial\svec/\partial\Psi\right|$ is the Jacobian of the
coordinate transformation. We can generate samples in
$[\mathbf{x}_0,\mathbf{x}_1]$ and evaluate
Eq. \ref{eq:marginal_likelihood_analytical} by Monte Carlo integration
using a reference distribution $\pi_s(\Psi)$, or
equivalently $\pi_s(\svec)=\pi_s(\Psi)|\partial\Psi/\partial\svec|$:
\begin{equation}
	\label{eq:marginal_likelihood_integral_monte_carlo}
	{\cal Z}(\pvec) \approx \left<
	\frac{
		\mathcal{N}\left(\Psi|\mean,\cov\right)
		\comp\left(\svec(\Psi)|\pvec\right)
		\sqrt{g_\svec}
	}{
		\pi_s(\Psi\mid\mathbf{x}_0,\mathbf{x}_1)
	}
	\left|\frac{\partial\svec}{\partial\Psi}\right|
	\right>_{\svec'_i\sim\pi_s(\mathbf{x}_0,\mathbf{x}_1)}
\end{equation}
We improve performance by considering several choices for $\pi_s$.
For narrow populations, we choose
$\pi_s(\lambda)\propto\comp(\lambda)$; for narrow event likelihoods,
we choose $\pi_s(\Psi)\propto{\cal N}(\Psi)$; and for broad
populations and event likelihoods, we sample uniformly. In practice,
to avoid hard-coding a choice based on the population
hyperparameters, we implement all three approaches in parallel and
retain the result with the largest effective Monte Carlo sample size.

\subsection{Synthetic Population}
\label{sec:sub:syn_population}

We draw the true compact-binary parameters, including mass, spin,
eccentricity, and redshift, from the desired distributions. We then
add measurement errors to generate synthetic PE samples using the
method described in Sec. III.A of \cite{2024PhRvD.110f3009Z}. The
complete procedure is summarized in Eq. \ref{eq:syn_pop_recipe}.
Boundary prescriptions for the synthetic tests are specified below;
Eq.~\ref{eq:syn_pop_recipe} summarizes the general error recipe:
\begin{equation}
	\label{eq:syn_pop_recipe}
	\begin{aligned}
		m_1,m_2         & \sim \prob(m_1,m_2)
		\\
		\chi_{1,z}      & \sim \mathcal{N}_{[-1,1]}(\mu_{\chi_1},
		\sigma_{\chi_1})
		\\
		\chi_{2,z}      & \sim \mathcal{N}_{[-1,1]}(\mu_{\chi_2},
		\sigma_{\chi_2})
		\\
		\mathcal{M}_c^i & = \mathcal{M}_c(m_1,m_2)
		\left[1+\beta(\tilde{r}_i+r_0)\right]
		\\
		\eta^i          & = \eta(m_1,m_2)
		\left[1+0.03(\tilde{r}'_i+r'_0)\frac{12}{\rho}\right]
		\\
		\chi_{1,z}^i    & \sim \chi_{1,z} + \tilde{\sigma}_{\chi_{1,z}}
		(r'_{\chi_{1,z}} + r_{\chi_{1,z},i}) \frac{12}{\rho}            \\
		\chi_{2,z}^i    & \sim \chi_{2,z} + \tilde{\sigma}_{\chi_{2,z}}
		(r'_{\chi_{2,z}} + r_{\chi_{2,z},i}) \frac{12}{\rho}
	\end{aligned}
\end{equation}
Here, ${\cal N}_{[-1,1]}(\mu,\sigma)$ denotes a normal distribution
with location $\mu$ and scale $\sigma$, truncated to $[-1,1]$. All
$r_\alpha$ and $r'_\alpha$ are drawn from a standard normal
distribution; primed values are independent of the sample index and
are held fixed across all sample draws. Although this procedure omits
correlations found in real GW posteriors, particularly correlations
between the component spins, it captures the principal trends across
parameter space and with signal amplitude. In
Appendix \ref{ap:boundaries}, we discuss
the role of sampling boundary conditions on the
error model and resulting distributions. The spinning, eccentric BBH test in Sec.~\ref{sec:full_model}
uses spin and eccentricity normals centered
on the injected values, with width $0.1$, truncated to $[-1,1]$ and
$[0,1]$, respectively; these draws omit the shared offsets and SNR
rescaling in the general recipe. Unphysical mass draws are discarded.

In the algorithm described by Eq. \ref{eq:syn_pop_recipe}, the
parameters $m_1$, $m_2$, $\chi_{1,z}$, and $\chi_{2,z}$ represent the
true component masses and aligned spin components, respectively. The
quantities $\mathcal{M}_c^i$, $\eta^i$, $\chi_{1,z}^i$, and
$\chi_{2,z}^i$ represent the $i$th synthetic posterior sample of chirp
mass, symmetric mass ratio, and the primary and secondary aligned spin
components, respectively. The parameters
$\tilde{\sigma}_{\mathcal{M}_c}$, $\tilde{\sigma}_{\eta}$,
$\tilde{\sigma}_{\chi_{1,z}}$ and $\tilde{\sigma}_{\chi_{2,z}}$ control the
measurement-error distribution for each event. In summary, we draw
true values from the desired distributions, convert the component
masses to $\mathcal{M}_c$ and $\eta$ using
Eqs. \ref{eq:chirp_mass} and \ref{eq:symmetric_mass_ratio}, add
realistic measurement errors, and generate $N_{\text{samples}}$
synthetic posterior samples for each event. Finally, we convert the
chirp-mass and symmetric-mass-ratio samples back to component masses
using Eqs. \ref{eq:inverse_chirp_mass} and
\ref{eq:inverse_symmetric_mass_ratio}.

\begin{align}\label{eq:chirp_mass}
	\mathcal{M}_c & = \frac{(m_1 m_2)^{3/5}}{(m_1 + m_2)^{1/5}},
	\\ \label{eq:symmetric_mass_ratio}
	\eta ~        & = \frac{m_1 m_2}{(m_1 + m_2)^2},
	\\
	\label{eq:inverse_chirp_mass}
	m_1           & = \frac{1}{2}\mathcal{M}_c \eta^{-3/5}
	\left(1+\sqrt{1-4\eta}\right),
	\\\label{eq:inverse_symmetric_mass_ratio}
	m_2           & = \frac{1}{2}\mathcal{M}_c \eta^{-3/5}
	\left(1-\sqrt{1-4\eta}\right).
\end{align}

For the synthetic PE tests, the reference-prior density is
$\pi_{\rm ref}(\lambda)=C\,\mathbf{1}_{\mathcal D}(\lambda)$ with
respect to $\mathrm{d}m_1\,\mathrm{d}m_2\,\mathrm{d}\chi_{1z}\,
\mathrm{d}\chi_{2z}\,\mathrm{d}\epsilon$, on the adopted physical
domain $\mathcal D$. This is the mock prior, rather than a particular
LVK event-analysis prior; its constant normalization cancels in the
population comparison. We request 5000 draws per event in the full-model test;
removing unphysical mass draws leaves 585--5000 draws for each of its
74 events. As demonstrated below, our
synthetic-data procedure performs well in end-to-end population
tests. The method is less successful in extreme regimes, such as
those involving broad spin distributions and large measurement
errors. For simplicity, we do not investigate scenarios in which our
recovery has been shown to produce biased population-parameter
estimates.

\subsection{Continuous likelihood representations}
Several strategies can be used to construct a continuous likelihood
model from either the underlying analysis components or posterior
samples produced by an upstream parameter-estimation method.

First, posterior samples can be converted into a continuous
approximation using density estimation. The simplest model is a
single multivariate Gaussian whose mean and covariance are estimated
from the posterior samples described above. Because of complications
associated with coordinate systems, boundaries, and priors, a
(truncated) multivariate Gaussian approximation is effective only in
suitable coordinates $y$ (e.g., $\mathcal{M}_c,\eta,\ldots$). Our
objective is therefore to approximate the posterior density
$p(x)dx=p(x(y))J(y(x))dy\equiv p(y)dy$ with a Gaussian. The
expression $p(y)/J(y)$ estimates the posterior in the original
coordinates; we then divide by the prior density so that
$\ell(x)=p(y)/[J(y)\pi(x)]$. In terms of $y$, we estimate the parameters
of the (truncated) multivariate Gaussian using conventional
estimators for an untruncated distribution:
\begin{align}
	\boldsymbol{\mu}_{\mathrm{y}}    & = \frac{1}{N_{\text{samples}}}
	\sum_{i=1}^{N_{\text{samples}}} \mathrm{y}^i,
	\\
	\boldsymbol{\Sigma}_{\mathrm{y}} & = \frac{1}{N_{\text{samples}}}
	\sum_{i=1}^{N_{\text{samples}}} (\mathrm{y}^i -
	\boldsymbol{\mu})(\mathrm{y}^i - \boldsymbol{\mu})^\top.
\end{align}

Second, the likelihood may be estimated directly using an alternative
method derived from the underlying inference strategy. For synthetic
PE samples, for example, the true mean and covariance are known, so
we can use this model directly as the likelihood. This choice
slightly changes the synthetic problem because the synthetic PE
procedure supplies only a likelihood, does not explicitly incorporate
the reference prior $\pi_{\rm ref}$, and ignores boundary effects.
For real data, some inference or analysis methods can provide a
direct likelihood estimate. For example, RIFT provides a likelihood
marginalized over extrinsic parameters. In addition,
Delfavero et al.
\cite{2022arXiv220514154D,2022APS..APRX16004D,2021arXiv210713082D}
demonstrated that Gaussian approximations accurately represent the
multidimensional likelihoods of real GW observations and capture many
of their correlated features. That study also provides Gaussian
approximations for nonprecessing and precessing parameter inferences
of GWTC-3 observations
\cite{2022arXiv220514154D,2022APS..APRX16004D,2021arXiv210713082D}.
Our code, \gwk, can use these Gaussian approximations directly and,
in the data-dominated limit, generate synthetic posterior samples of
any required size. These phenomenological Gaussians therefore allow
us to characterize directly how finite-sample effects influence
realistic astrophysical conclusions.

Our companion RIFT-distance-likelihood study (hereafter RDL)
\cite{hemrattaphanLikelihoodReconstruction} develops a framework
for continuous likelihood approximations. It converts reported
posterior samples into continuous estimates, accounts for their
reference priors, and supplements that information with
distance-resolved RIFT likelihood evaluations. The reconstructed
likelihoods are validated by recovering the full reference posterior,
including its correlations and its response to population reweighting.
The public data release and repository provide fitted models and
their validation records~\cite{rdlLikelihoodRelease,rdlLikelihoodRepository}.
Because this framework is more involved and developed for real
observations, we use it in the real-event analyses below. The main
synthetic-population tests use Gaussian likelihood models.

RDL addresses fit-quality systematics, including non-Gaussian
structure and physical boundaries. Truncation-aware density estimation
also offers methods for constructing these approximations
\cite{melchiorGoulding2018}. Here we focus on the numerical integration
of a specified continuous likelihood.

\begin{figure*}[t]
\centering
\includegraphics[width=0.48\textwidth]{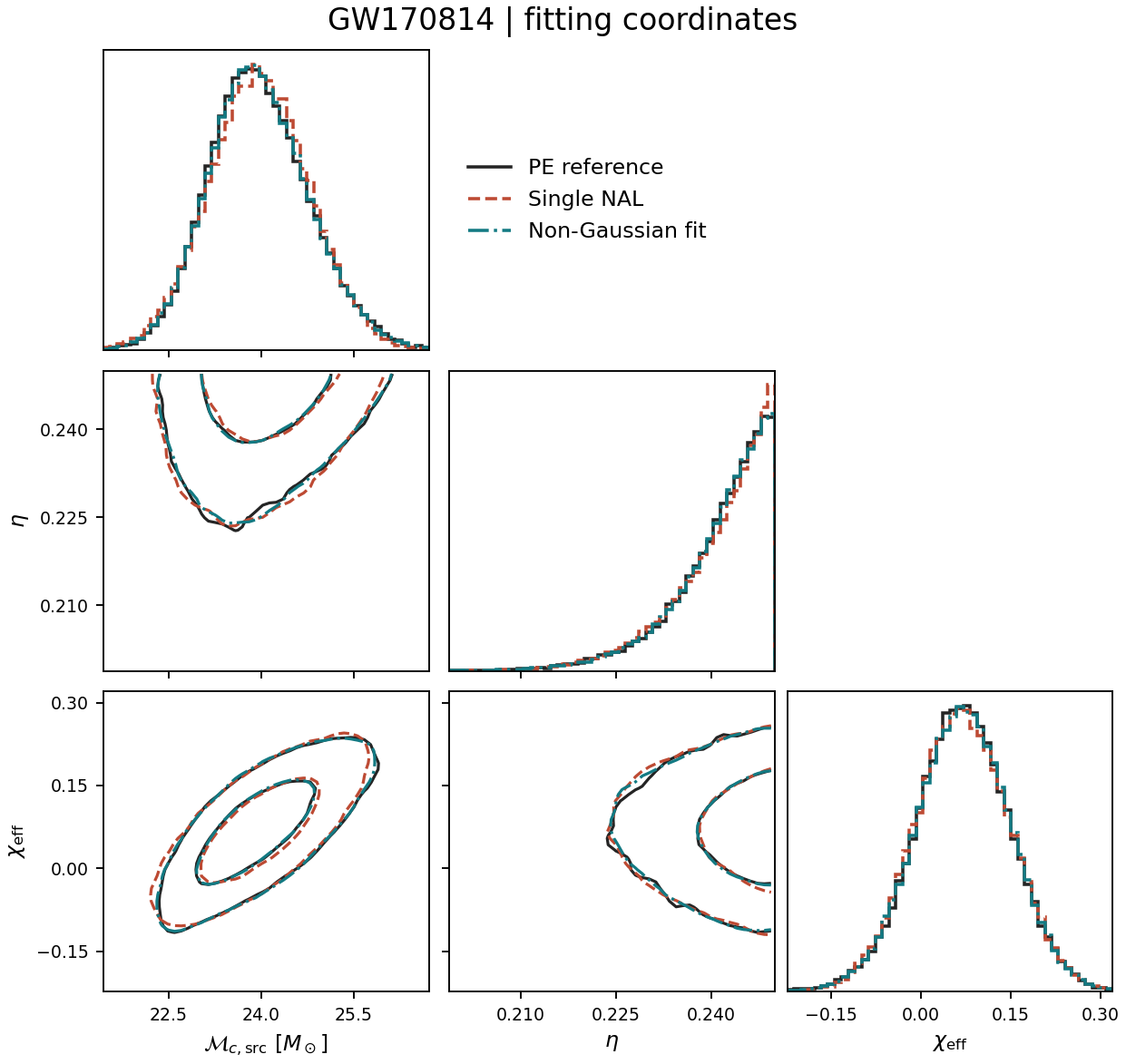}
\includegraphics[width=0.48\textwidth]{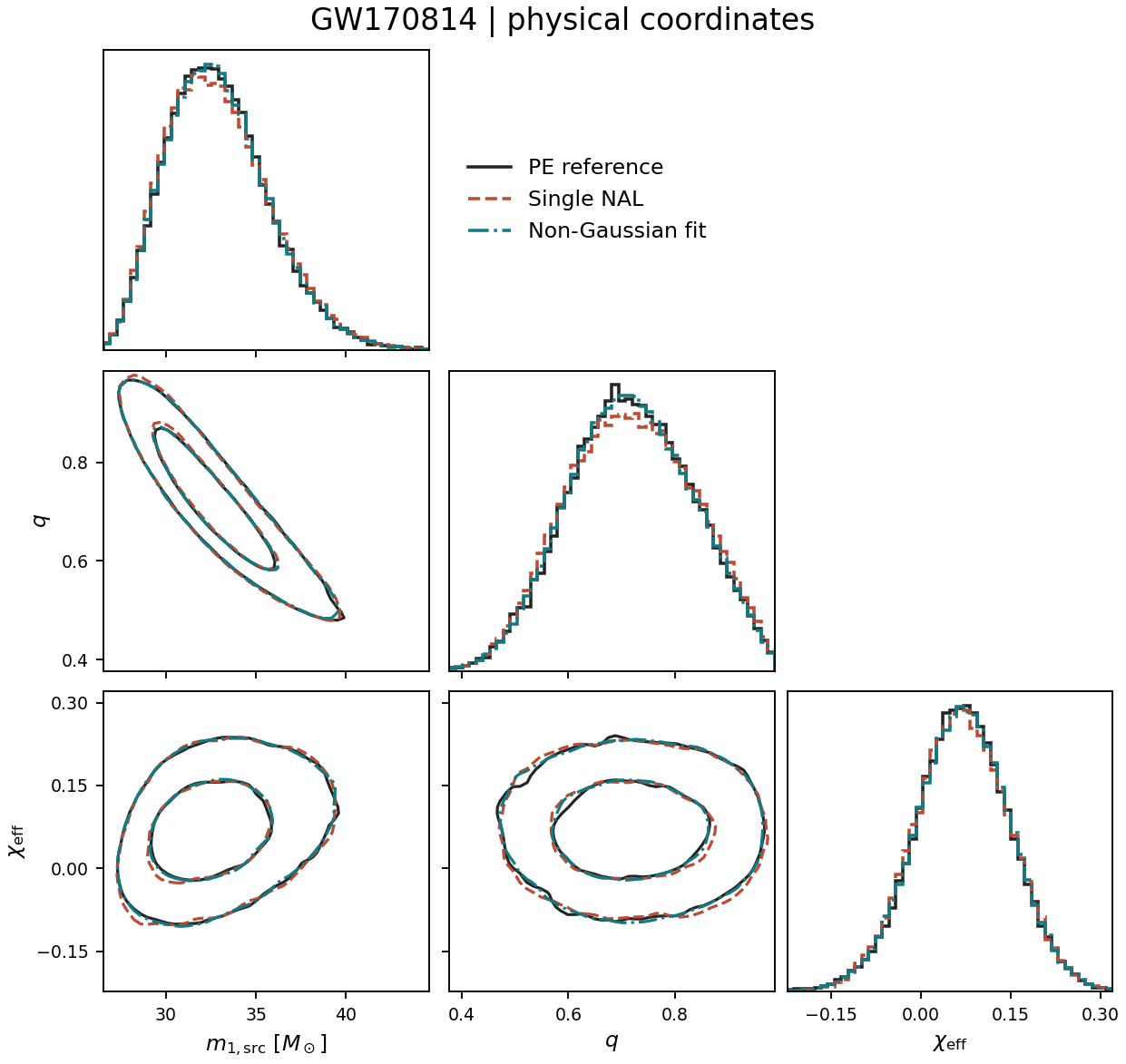}
\caption{GW170814 under the common comparison prior, in fitting
coordinates (left) and physical coordinates (right). The PE reference,
Delfavero et al.\ NAL fit, and non-Gaussian RDL fit agree well in this example.
Redshift has been marginalized; both panels represent the same
three-dimensional distribution. \llmfig}
\label{fig:good_event_fit}
\end{figure*}

To illustrate the efficacy of these representations,
Fig.~\ref{fig:good_event_fit} compares the NAL fit provided by
Delfavero et al.~\cite{2021arXiv210713082D} with a non-Gaussian RDL
fit for GW170814 and the PE reference. We use \emph{base8} below
as shorthand for the RDL fit's eight-component internal
representation. The comparison uses a
common prior flat in $(\mathcal M_{c,\rm src},\eta,\chi_{\rm eff})$,
after marginalizing redshift with the source-time comoving-volume
measure. The right panel expresses the same distributions in
physical coordinates while retaining that comparison prior. The PE
reference is reweighted to this comparison prior, isolating
likelihood-shape differences under a common measure.

We used OpenAI Codex (GPT-6.1 Sol) to assist with implementing and running
the real-event comparison and its numerical checks
(Appendix~\ref{app:llm}).

\section{Analyses}\label{sec:analyses}
Black-hole subpopulations spanning a narrow range of properties are
both astrophysically plausible and practically useful. For example,
populations with well-determined masses could serve as standard
sirens for cosmology. In addition, some studies of rotating massive
stars suggest that their cores rotate very slowly, naturally
producing black holes with nearly zero spin. In this section, we
analyze several synthetic populations. We begin with a toy model in
which the mass-distribution parameters are fixed and only the aligned
spin components are inferred, thereby illustrating the advantages of
the continuous-likelihood approach and the limitations of the
discrete approach. We then compare the two approaches using a full
model that includes mass, spin, and eccentricity. Finally, we extend
the analysis to a multisource model containing BNS, BBH, and NSBH
populations with different characteristic spin scales.

\subsection{Proof of Concept with Toy Population Model}

\begin{figure*}
	\centering

	\includegraphics[width=0.45\textwidth]{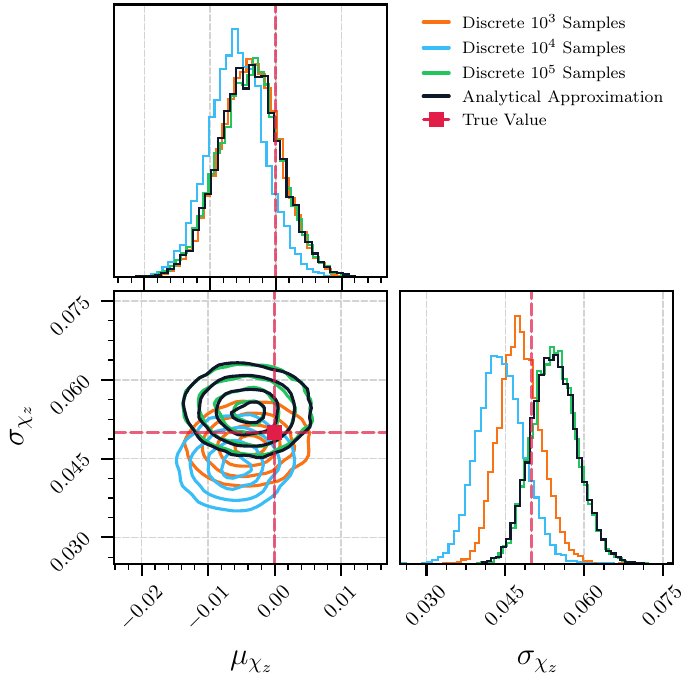}
	\includegraphics[width=0.45\textwidth]{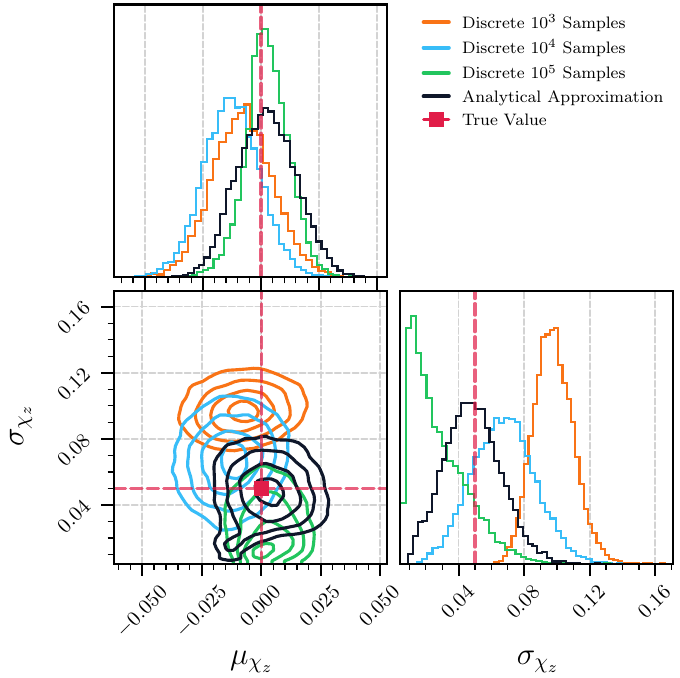}
	\includegraphics[width=0.45\textwidth]{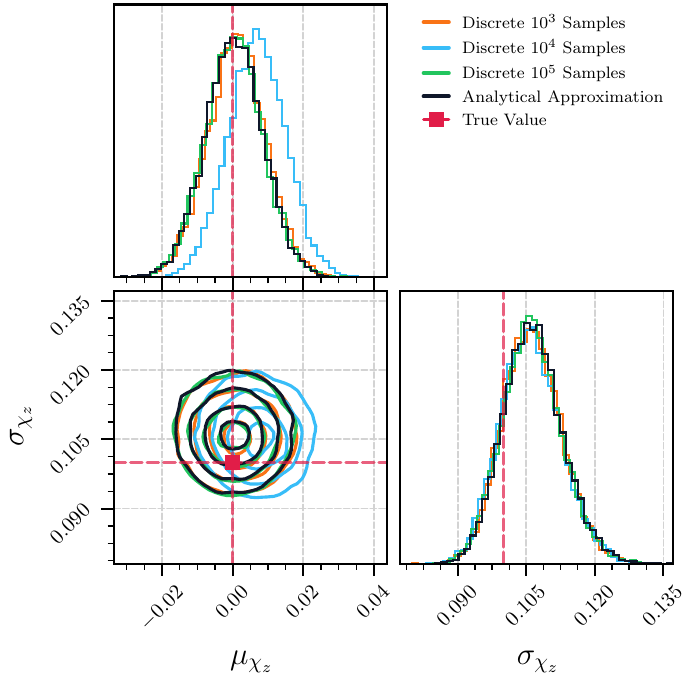}
	\includegraphics[width=0.45\textwidth]{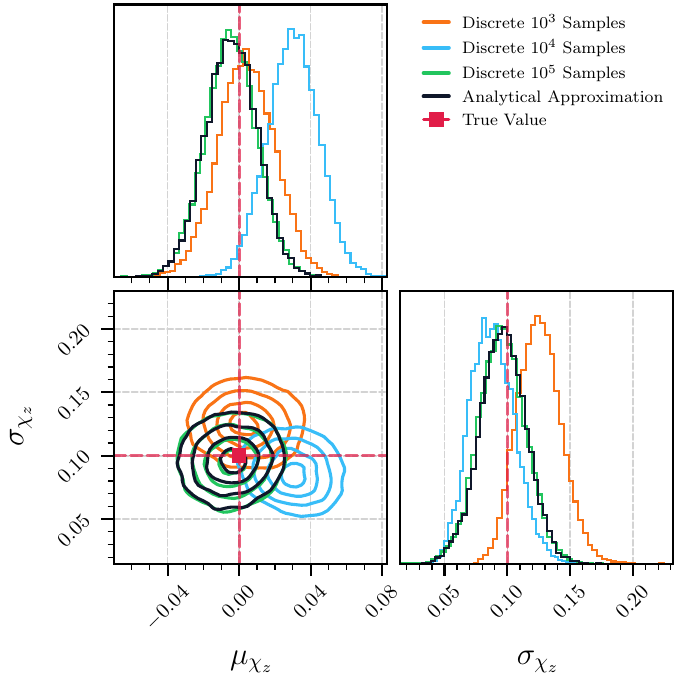}

	\caption{Recovery of synthetic spin distributions from populations of synthetic events using
		different single-event characterization strategies. Only the two spin distribution parameters
		are inferred, assuming a known mass distribution (see Table \ref{tab:toy_model_parameters}).
		Colors distinguish the characterization strategies: discrete (varying sample sizes) or a
		continuous approximation (fit to $10^5$ samples). Rows correspond to population sizes:
		115 events (top) and 104 events (bottom). Columns correspond to single-event measurement uncertainties:
		$\tilde{\sigma}_{\chi_{z}}=0.05$ (left) and $\tilde{\sigma}_{\chi_{z}}=0.2$ (right);
		see Sec. \ref{sec:sub:syn_population} for details on $\tilde{\sigma}_{\chi_z}$.}
	\label{fig:corner_plot_toy_model}
\end{figure*}

Figure \ref{fig:corner_plot_toy_model} shows the recovery of a
synthetic population using four different representations of the
single-event likelihoods. In these calculations, we fix all
population parameters except the two parameters that characterize the
spin distribution. The fixed parameters match the known generative
model. We choose population parameters that produce reliable
synthetic data while avoiding subtle boundary effects. The four
calculations differ only in how they evaluate the
single-event marginal likelihood integrals ${\cal Z}_j$
[Eq. \ref{eq:marginal_likelihood_approximation}]. In the first three
calculations, we perform inference using the
discrete method with three different sample sizes ($10^3$,  $10^4$,
$10^5$); in all cases, samples are drawn directly from the
synthetic single-event posteriors
[Eq. \ref{eq:marginal_likelihood_discrete_standard}]. In the final calculation,
we instead perform these integrals using Monte Carlo integration with
well-tuned normal sampling distributions for each event
[Eq. \ref{eq:marginal_likelihood_integral_monte_carlo}]. The
continuous method yields reliable estimates whether the intrinsic
population scale is larger or smaller than the measurement-error
scale. The discrete and continuous approaches produce consistent
results when sufficiently many samples are available and the
measurement-error scale is smaller than the intrinsic population
scale, as shown in the left column of
Fig. \ref{fig:corner_plot_toy_model}. When fewer samples are
available or the measurement-error scale exceeds the intrinsic
population scale, the discrete approach struggles to recover the true
parameters, whereas the continuous approach remains reliable, as
shown in the right column.

\begin{table}[t]
	\centering
	\begin{ruledtabular}
		\begin{tabular}{lcc}
			Parameters            & Synthetic Population & Prior Range           \\\hline
			$\ln(\rate)$          & 3.3                  & Fixed                 \\
			$\mu_{m_1}$           & 40.0                 & Fixed                 \\
			$\mu_{m_2}$           & 20.0                 & Fixed                 \\
			$\sigma_{m_1}$        & 6.0                  & Fixed                 \\
			$\sigma_{m_2}$        & 3.0                  & Fixed                 \\
			$\mu_{\chi_{1,z}}$    & 0.0                  & $[-1,1]$              \\
			$\mu_{\chi_{2,z}}$    & 0.0                  & $\mu_{\chi_{1,z}}$    \\
			$\sigma_{\chi_{1,z}}$ & \{0.05, 0.1\}        & $[0,1]$               \\
			$\sigma_{\chi_{2,z}}$ & \{0.05, 0.1\}        & $\sigma_{\chi_{1,z}}$
		\end{tabular}
	\end{ruledtabular}
	\caption{True parameters used to generate the synthetic populations
		and prior ranges for the toy population model. The true values of
		$\sigma_{\chi_{1,z}}$ and $\sigma_{\chi_{2,z}}$ correspond to two
		separate analyses. We fix the mass-distribution parameters to focus
		on recovering the parameters of the aligned-spin distribution.}
	\label{tab:toy_model_parameters}
\end{table}

\subsection{Demonstration with Synthetic Population using Full Model}
\label{sec:full_model}

\begin{figure*}[t]
	\centering
	\includegraphics[width=\textwidth]{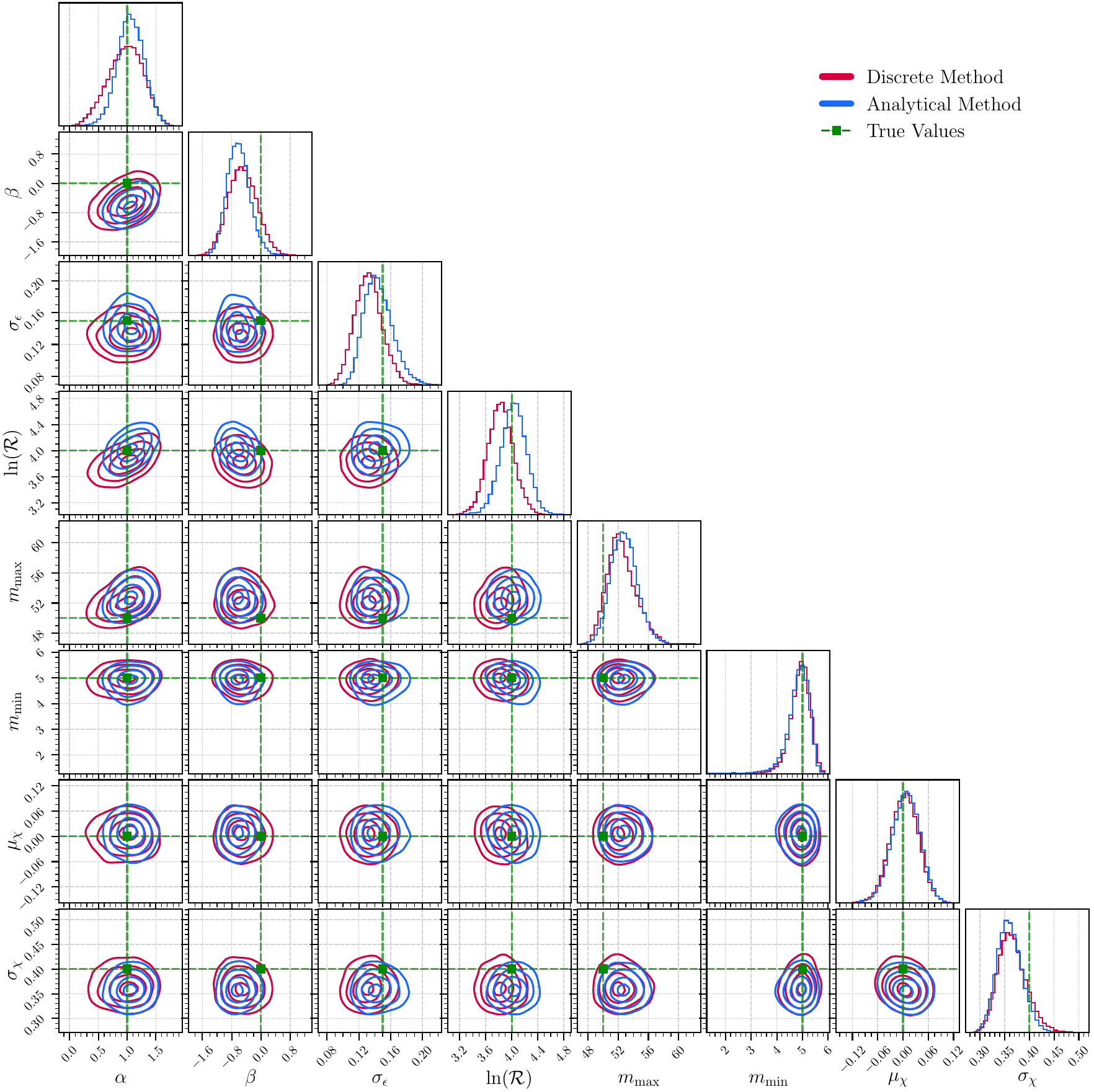}
	\caption{\label{fig:result_b_ecc_spin}Recovery of the full synthetic
		population using discrete and continuous approximations to the
		single-event likelihoods. The continuous approximation captures
		the inferred mass, spin, eccentricity, and rate parameters.}
\end{figure*}

Figure \ref{fig:result_b_ecc_spin} shows the recovery of a synthetic
population of binary sources, each characterized by two masses, two
aligned spin components, and an orbital eccentricity. We again
compare results obtained using discrete and continuous
approximations. In this BBH analysis, the distribution of aligned
spin components is modeled by a \textsc{Truncated Gaussian}, and the
component masses are described by the \textsc{Powerlaw Primary +
MassRatio} model:

\begin{equation}
	\begin{aligned}
		p(m_1|\pvec_0)       & \propto m_1^{-\alpha}, \quad m_{\min}
		\leq m_1 \leq m_{\max}                                        \\
		p(q|m_1, \pvec_1)    & \propto q^{\beta}, \quad m_{\min} \leq
		m_1q \leq m_1                                                 \\
		p(m_1,m_2\mid \pvec) & = p(m_1|\pvec_0) p(q|m_1, \pvec_1)/m_1
	\end{aligned}
\end{equation}

where $m_1$ and $m_2$ are the primary and secondary masses,
respectively; $q=m_2/m_1$ is the mass ratio;
$\pvec_0=\{\alpha,m_{\min},m_{\max}\}$;
$\pvec_1=\{\beta,m_{\min}\}$; and
$\pvec=\pvec_0\cup\pvec_1$ is the set of population hyperparameters.
Table \ref{tab:full_model_parameters} provides the full set of
hyperparameters used to generate the synthetic population and the
corresponding hyperparameter prior ranges.
After selecting the hyperparameters and drawing synthetic events from
the model, we generate synthetic PE samples using the procedure
described in Sec. \ref{sec:sub:syn_population}, with
$\tilde{\sigma}_{\mathcal{M}_c}=\tilde{\sigma}_{\eta}=1.0$,
$\tilde{\sigma}_{\chi_{1z}}=\tilde{\sigma}_{\chi_{2z}}=
\tilde{\sigma}_{\epsilon}=0.1$. We request 5000 samples per event
before discarding unphysical mass draws.
Further details are provided in Sec. III.B of \cite{krnm-3vrf}.

The figure demonstrates that continuous likelihoods can be used in a
full population-inference analysis to recover the injected
population without introducing an apparent bias.

\begin{table}[t]
	\centering
	\begin{ruledtabular}
		\begin{tabular}{lcc}
			Parameters            & Synthetic Population & Prior Range           \\\hline
			$\ln(\rate)$          & 4.0                  & $[0,10]$              \\
			$\alpha$              & 1.0                  & $[-6,6]$              \\
			$\beta$               & 0.0                  & $[-6,6]$              \\
			$m_{\min}$            & 5.0                  & $[1,20]$              \\
			$m_{\max}$            & 50.0                 & $[30,80]$             \\
			$\mu_{\chi_{1,z}}$    & 0.0                  & $[-1,1]$              \\
			$\mu_{\chi_{2,z}}$    & 0.0                  & $\mu_{\chi_{1,z}}$    \\
			$\sigma_{\chi_{1,z}}$ & 0.4                  & $[0,1]$               \\
			$\sigma_{\chi_{2,z}}$ & 0.4                  & $\sigma_{\chi_{1,z}}$ \\
			$\mu_{\epsilon}$      & 0.0                  & Fixed                 \\
			$\sigma_{\epsilon}$   & 0.15                 & $[0,1]$
		\end{tabular}
	\end{ruledtabular}
		\caption{True parameters used to generate the synthetic population
			and prior ranges used for the full population model.}
	\label{tab:full_model_parameters}
\end{table}

\subsection{Demonstration with a Synthetic Population Using a Multisource Model}
\begin{figure*}[t]
	\centering
	\includegraphics[width=0.95\textwidth]{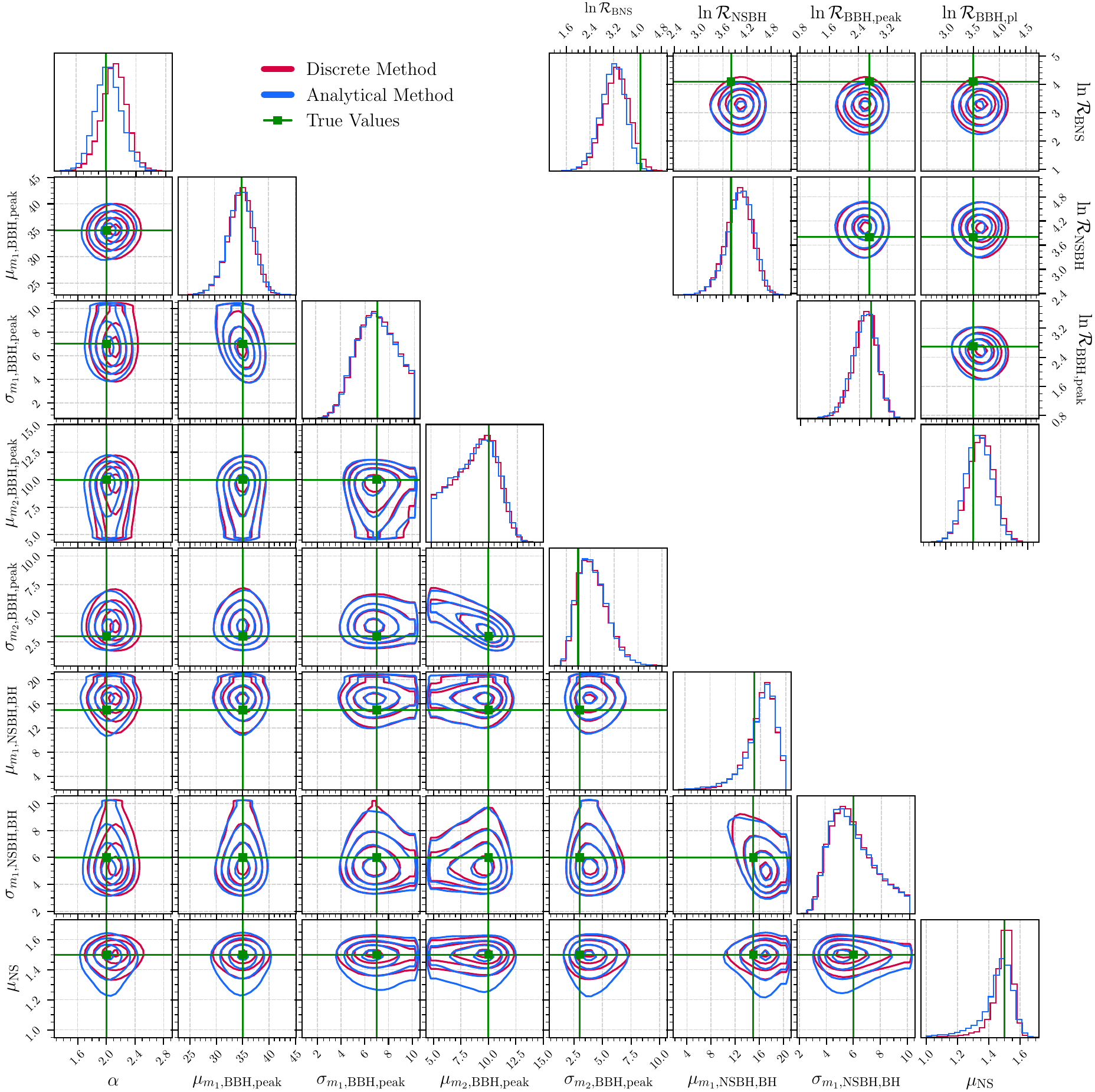}

	\caption{Population mass distribution and merger rate parameter recovery
		for the multisource model. Because the parameters do not occupy a
		narrow region, the continuous and discrete approaches yield similar
		recoveries. See Fig. \ref{fig:multi_source_spin.pdf} for the spin
		distribution parameters.}

	\label{fig:multi_source_mass_rate.pdf}
\end{figure*}

\begin{figure}[t]
	\centering
	\includegraphics[width=0.95\columnwidth]{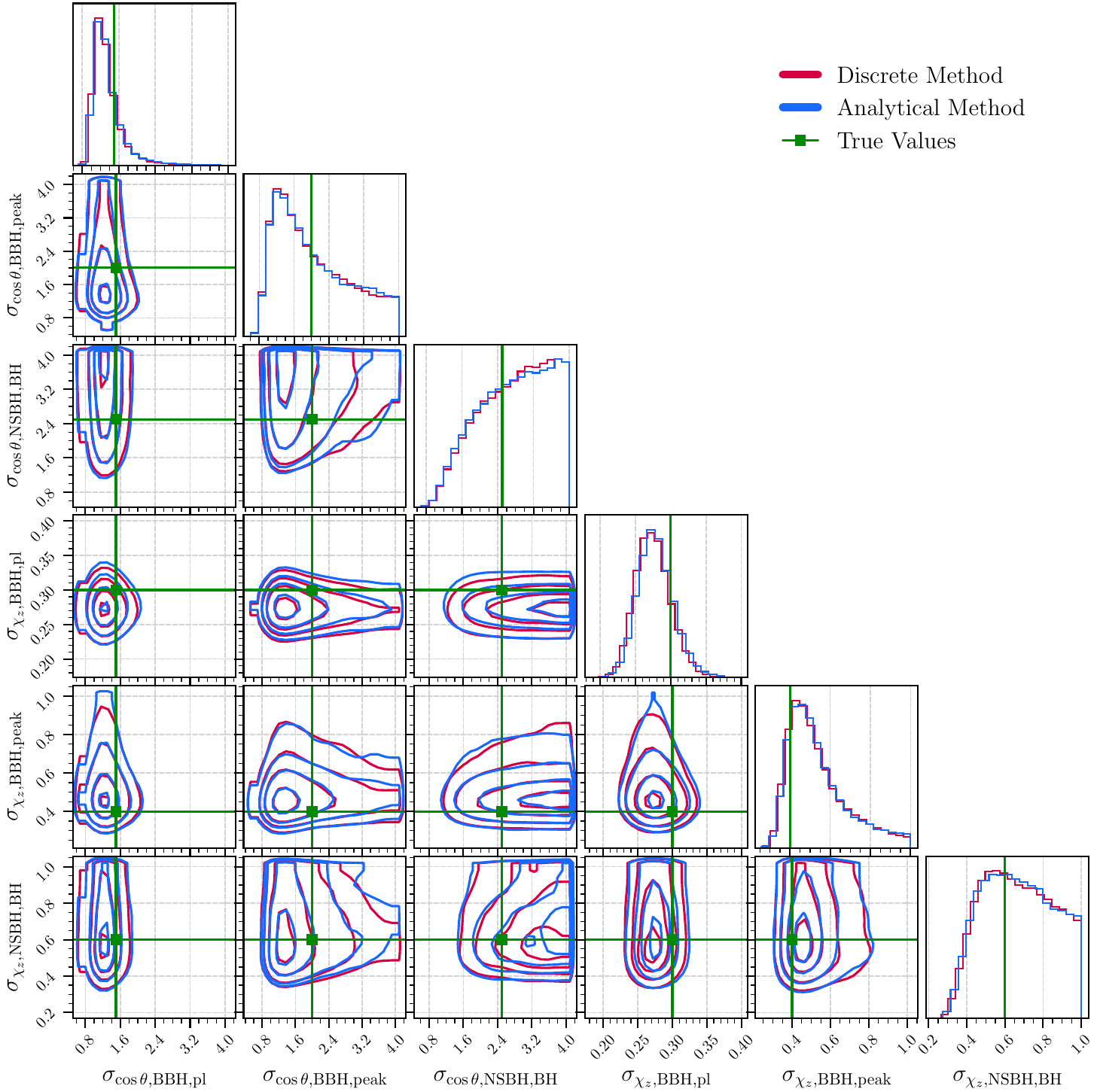}

	\caption{Recovery of the spin magnitude and cosine tilt distribution
		parameters for the multisource model. As with the mass parameters,
		these parameters do not occupy a narrow region; consequently, the
		continuous and discrete approaches yield consistent results.}
	\label{fig:multi_source_spin.pdf}
\end{figure}

Figures \ref{fig:multi_source_mass_rate.pdf} and
\ref{fig:multi_source_spin.pdf} show results obtained from another
synthetic data set, generated from a multipopulation model containing
BNS, NSBH, and BBH sources. The model is defined in previous work
\cite{krnm-3vrf} and closely resembles the multisource framework used
to interpret GWTC-3 \cite{physrevx.13.011048}. We use the same set of
synthetic events generated for that earlier study \cite{krnm-3vrf},
but regenerate the synthetic posterior samples for each event using
the procedure described above.
These figures again demonstrate that continuous likelihoods can be
used in a full population-inference analysis to recover an injected
population with multiple components without introducing an apparent
bias.

\subsection{Recovery of the GWTC-4.0 low-mass peak with real events}
\label{sec:real_event_recovery}

We test whether continuous event likelihoods recover the low-mass
feature in the GWTC-4.0 binary black hole population
\cite{2025arXiv250818083T}. We compare three analyses of the same 153
events from the companion study~\cite{hemrattaphanLikelihoodReconstruction},
selected with false-alarm rate below $1\,\mathrm{yr}^{-1}$.
The all-PE analysis resamples the matched PE reference for every
event. The primary RDL analysis uses the adopted non-Gaussian
likelihoods for all 153 events, including GW150914. A third,
restricted-fit analysis substitutes 67 fits that match the public
release and pass the validation criteria described below. It uses
the paired PE references for
the other 86 events. The primary RDL analysis requires the native
shape checks on each fit's own reference posterior. The fits retain
the two masses, effective aligned spin, and redshift. The three
analyses use a common selection model evaluated with the GWTC-4.0 search
injections~\cite{gwtc4SensitivityRelease}. Appendix~\ref{ap:VT}
describes the selection calculation.

The holdout tests assess how well a fit represents posterior samples
withheld from its training. Agreement with the training samples alone
can conceal errors in tails or correlations that become important
when a population model reweights the likelihood. The restricted-fit
analysis therefore requires agreement with held-out reference samples
under the native prior and six specified population reweightings.
Comparing it with the enlarged fit set tests how much these
eligibility requirements change the recovered population. Appendix~\ref{ap:fit_coverage}
defines the sample splits and numerical criteria.

We adopt the GWTC-4.0 Broken Power Law + Two Peaks mass family
\cite{2025arXiv250818083T}. We vary the location $\mu_1$ and width
$\sigma_1$ of the first Gaussian peak and the three mixture weights
before applying the low-mass taper. We fix the broken power law,
tapers, mass-ratio law, second Gaussian, and redshift evolution at
their released marginal medians~\cite{gwtc4PopulationRelease}. All
three analyses use a fixed Gaussian effective-spin distribution with mean
0.04 and standard deviation 0.11. The event and selection integrals
cover $5\,M_\odot \leq m_2 \leq m_1 \leq 300\,M_\odot$.
The RDL inputs retain only effective aligned spin; we impose the
$5\,M_\odot$ component-mass floor in both integrals. Reproducing the full
GWTC-4.0 population posterior would require the component-spin
likelihoods and matching event and selection inputs, including
support for secondary masses below $5\,M_\odot$.

\begin{figure*}[t]
	\centering
	\includegraphics[width=0.98\textwidth]{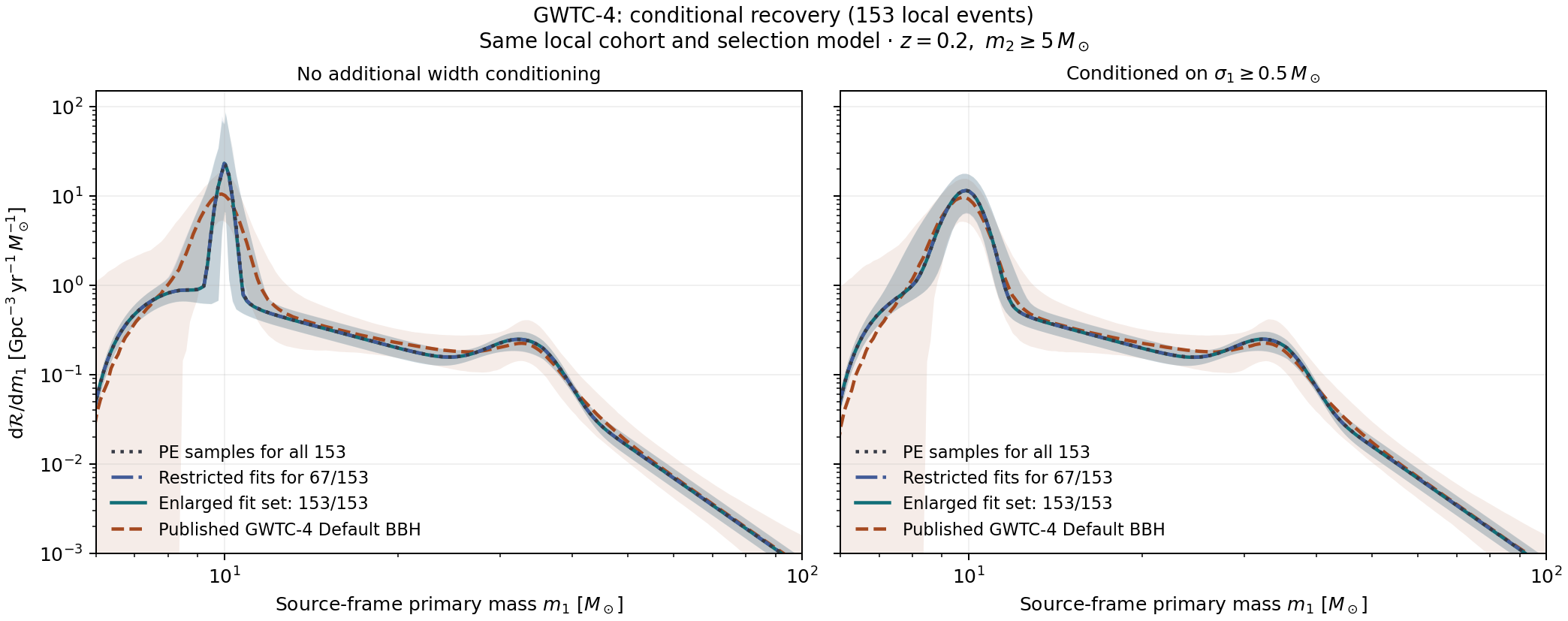}
	\caption{Conditional recovery of the GWTC-4.0 first primary-mass
		peak with PE samples for all 153 events (gray dotted), restricted
		RDL fits for 67 events (blue dash-dot), and adopted RDL fits for
		all 153 events (solid teal). The restricted analysis uses PE for
		the remaining 86 events. The GWTC-4.0 population release is
		shown in dashed rust. Curves and bands show
		the median and pointwise 90\% credible interval at $z=0.2$,
		integrating over $m_2\geq5\,M_\odot$. The left panel applies no
		additional width conditioning; the local width prior is
		$0.05\leq\sigma_1/M_\odot\leq10$. The right panel conditions
		all four results on $\sigma_1\geq0.5\,M_\odot$. Local analyses
		use fixed-injection selection and no additional likelihood-variance
		cut. Published draws retain their native prior and catalog cut.
		The narrow-width results remain sensitive to integration banks
		and grid refinement, as quantified in the text. \llmfig}
	\label{fig:gwtc4_mass_recovery}
\end{figure*}

All three analyses place the first peak at a median of $10.08\,M_\odot$,
close to the published median of $9.76\,M_\odot$. Figure
\ref{fig:gwtc4_mass_recovery} shows both the full-width and
$\sigma_1\geq0.5\,M_\odot$ comparisons of the differential merger rate
$\mathrm{d}\mathcal{R}/\mathrm{d}m_1$ at $z=0.2$. For each shape
draw we sample the rate amplitude conditional on the selection
integral under a log-uniform rate prior. We apply the same
$m_2\geq5\,M_\odot$ cut to the published rate draws. With all tested
widths, the peak height is sensitive to the integration banks.
Repeating the fit with two disjoint halves of the event and selection
banks yields median rates of 38.7 and 16.6 at $m_1=10\,M_\odot$ for
the analysis using RDL likelihoods, in units of
$\mathrm{Gpc}^{-3}\,\mathrm{yr}^{-1}\,M_\odot^{-1}$.
The PE analysis shows similar sensitivity. On the finer numerical
grid, about 21\% of the RDL analysis posterior lies in the narrowest
width cell. The peak width and height are unresolved in this regime.

Conditioning all analyses on $\sigma_1\geq0.5\,M_\odot$ gives
closer agreement between bank halves. In this slice the two RDL
half-bank medians are 11.6 and 11.2 at $10\,M_\odot$, and the
full-bank 90\% intervals are $6.4$--$17.5$ for the analysis using
RDL likelihoods and $6.4$--$17.6$ for the PE analysis. The published
interval is $5.0$--$15.3$ under the same cuts. The three local
full-bank medians at $10\,M_\odot$ are 11.35, 11.35, and 11.30
for the all-PE, restricted-fit, and primary RDL analyses,
respectively. The uncertainty away from the first peak is
conditional on the fixed mass parameters.

\paragraph{Effect of the likelihood-variance cut.}
We apply the GWTC-4.0 cut on the estimated Monte Carlo variance of
the rate-marginalized log likelihood, $\widehat{\mathrm{Var}}
[\ln\widehat{\mathcal{L}}]<1$~\cite{2025arXiv250818083T}, to the
same likelihood grid in the all-PE and primary RDL analyses. For PE
resamples the event
variance is conditional on the empirical posterior and excludes
uncertainty from the finite original PE sample. Within
$\sigma_1<0.5\,M_\odot$, the cut removes 29--34\% of the posterior
probability on the coarse grid and 40--46\% on the finer grid,
which extends to smaller widths. On the finer grid the cut removes
less than 1\% at larger widths. In the RDL analysis the median rate at
$10\,M_\odot$ falls from 23.2 to 18.5, and the 95th percentile
falls from 64.6 to 38.4. The injection term accounts for most
of the removal, and the event term alone leaves essentially all
of the posterior probability in both analyses. The older GWTC-3.0
effective-sample-size cuts also leave essentially all of the probability.
The cut therefore preferentially excludes narrow, high peaks in this
test. The uncut bank disagreement leaves the underlying peak width
unresolved, and replacing the event likelihoods alone leaves the
shared selection uncertainty in place.

\subsection{Extension to the GWTC-5 catalog}
\label{sec:gwtc5_recovery}

We repeat the comparison after adding the 104 O4b BBHs listed in the
GWTC-5 population release~\cite{gwtc5PopulationRelease}. The resulting
cohort has 257 events, including GW150914. We retain the earlier
PE sources and nuisance parameters and use the cumulative GWTC-5
selection injections~\cite{gwtc5SensitivityRelease}. These choices
isolate the effect of extending the catalog. The published GWTC-5
analysis uses 259 BBHs and differs in some reference-waveform choices~\cite{gwtc5PopulationPaper}.
The RDL analysis adds all 97 available four-dimensional long-EM fits
passing the native-posterior shape checks on three splits, giving 250
fit substitutions in total. The remaining seven events use PE
samples. The restricted analysis retains the 67 earlier fits and
adds 67 O4b fits that pass the full diagnostics on all three
validation splits, for 134 fit substitutions and 123 PE references.
Appendix~\ref{ap:fit_coverage} identifies the low-mass fallbacks and
reports a separate check with fits for all 257 events.

\begin{figure*}[t]
\centering
\includegraphics[width=0.98\textwidth]{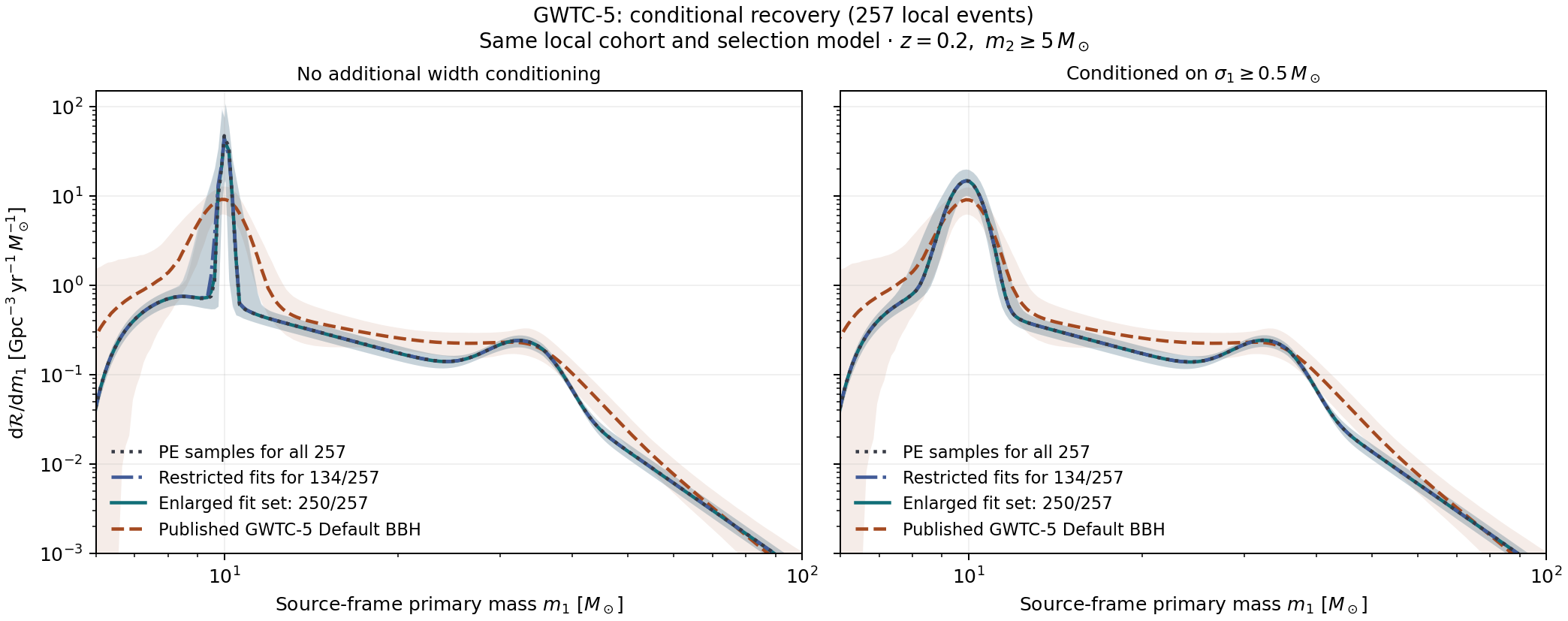}
\caption{GWTC-5 extension, with the same plotting conventions as
Fig.~\ref{fig:gwtc4_mass_recovery}. The left panel applies no
additional width conditioning; the right conditions the local and
published results on $\sigma_1\geq0.5\,M_\odot$. The local analyses
use the same 257 events, fixed nuisance parameters, and fixed-injection
selection. The three local curves use PE for every event, restricted
fits for 134 events, and the enlarged set of 250 fits, respectively.
Events without a fit in each arm use their matched PE references.
The published analysis uses 259 events and retains its native prior
and catalog variance cut. \llmfig}
\label{fig:gwtc5_mass_recovery}
\end{figure*}

Figure~\ref{fig:gwtc5_mass_recovery} compares the extended catalog
with its published Default BBH result, which uses the same Broken
Power Law + Two Peaks mass family. We project each published draw
to $z=0.2$ and restrict the secondary mass to
$m_2\geq5\,M_\odot$. The resolved-width comparisons also apply
$\sigma_1\geq0.5\,M_\odot$ to the published draws.
All three local representations again place the low-mass peak near
$10.08\,M_\odot$. With $\sigma_1\geq0.5\,M_\odot$, the primary RDL/PE
half-bank median rates at $10\,M_\odot$ are 14.5 and 14.8, and the
full-bank 90\% interval is $10.0$--$19.7$ in
$\mathrm{Gpc}^{-3}\,\mathrm{yr}^{-1}\,M_\odot^{-1}$.
The PE interval is also $10.0$--$19.7$. The three local full-bank
medians in this width slice are 14.69, 14.62, and 14.66 for the
all-PE, restricted-fit, and primary RDL/PE analyses, respectively.
The published GWTC-5 median under the same plotted cuts is 9.0, with a 90\% interval
of $6.2$--$12.5$. With all tested widths, the hybrid half-bank
medians differ substantially, at 67.2 and 28.8.

The variance cut has a stronger effect in this extended catalog.
On the finer grid it retains 4.0\% of the RDL/PE posterior
probability and 1.2\% of the PE posterior probability. In the RDL/PE
analysis the median rate at $10\,M_\odot$ falls from 45.5 to 13.7.
The selection term alone retains 8.1\%, while the event term alone
retains essentially all of the probability. The catalog size enters
the selection contribution to the log-likelihood variance
quadratically. These results show the increasing importance of
selection precision as the catalog grows, while leaving the height
of the very narrow peak unresolved in both event representations.

\subsection{Continuous selection check}
\label{sec:continuous_selection}

We add a catalog-calibrated neural selection calculation while
retaining the fixed-injection results above as controls. After
marginalizing the fixed secondary-mass, effective-spin, and redshift
distributions, we fit a continuous effective volume as a function of
primary mass (Appendix~\ref{ap:VT}). Each trial mass component is
integrated with new population draws, including direct draws from
its proposed narrow Gaussian. On the finer grid, the hybrid
half-bank median rates at $10\,M_\odot$ change from 38.7 and 16.6
to 21.4 and 19.2 for GWTC-4, and from 67.2 and 28.8 to 28.5 and
27.4 for GWTC-5, in the units used above. This improves the observed
split-bank agreement. Nevertheless, 15.5\% and 14.5\% of the
respective full-bank neural posteriors occupy the narrowest width
cell, with a prior floor of $0.05\,M_\odot$. The width has no
resolved lower bound, and the peak height remains sensitive to grid
refinement. These conditional tests do not establish a bias in the
published, fully marginalized catalog inference. A direct comparison
of neural and injection-based recovery curves, with and without
the additional width condition, is shown in
Appendix~\ref{ap:selection_recovery}.

\section{Conclusion}\label{sec:conclude}

Using the continuous-likelihood method implemented in \gwk, we
recover the low-mass peak in the primary-mass distribution of merging
binary black holes in both GWTC-4 and GWTC-5. In these conditional
fits, the median peak location is $10.08\,M_\odot$ in each catalog and
agrees with the location recovered using posterior samples. The
inferred location is also qualitatively consistent with the
approximately $10\,M_\odot$ mass reported for the nearest black hole
newly identified in Gaia DR4 \cite{gaiaDr4DormantBlackHoles}. The two
observations probe a similar black-hole mass scale in different
binary populations.

The width and height of the peak are less well constrained than its
location. In both catalogs, uncertainty from the shared selection
injections dominates the effect of the likelihood-variance cut,
which preferentially removes narrower, higher peaks. A continuous
selection model improves agreement between independent halves of the
integration banks, but the width has no resolved lower bound and
the height remains sensitive to grid refinement. Measuring the
shape of the peak therefore requires better numerical control of
the selection and population integrals as well as reliable
per-event likelihoods.

Our synthetic-catalog tests show where continuous likelihoods make
the greatest difference. Both representations recover similar mass
distributions, while continuous likelihoods more accurately recover
narrow spin and eccentricity distributions. This matters when
assessing evidence for a population in which both the primary and
secondary black holes have very small spins. Only posterior draws
with both spin magnitudes in the narrow range contribute
appreciably to the population integral, and some events may have no
such draws. We estimate that resolving a population spin width
$\sigma_\chi$ requires $N_s>(\chi_{\rm max}/\sigma_\chi)^6$
posterior samples per event, where $\chi_{\rm max}$ is the
characteristic single-event spin measurement uncertainty. Halving
the width increases this sample requirement by a factor of 64.
A continuous likelihood allows us to evaluate the support for small
spins directly in that region, provided the likelihood fit is
accurate there.

Several approaches can supply continuous per-event likelihoods,
including the interpolation used by RIFT
\cite{gwastro-PopulationReconstruct-EOSStack-Wysocki2019}, kernel
density estimates \cite{2026RASTI...5ag012A}, Gaussian processes
\cite{2021MNRAS.508.2090D}, single Gaussians
\cite{2021arXiv210713082D}, Gaussian mixture models
\cite{2018MNRAS.479..601D}, normalizing flows
\cite{2026arXiv260614229A,2025PhRvD.111l3049M}, and neural networks
\cite{2025PhRvD.112l2003L}. As gravitational-wave catalogs grow,
combining vetted continuous likelihoods with precise selection
integrals will help determine how narrow features such as the
$10\,M_\odot$ peak are and what they reveal about compact-binary
formation.

\section*{Acknowledgements}
This material is based upon work supported by the NSF's LIGO
Laboratory, a major facility fully funded by the National Science
Foundation. The authors acknowledge the computational resources
provided by the LIGO Laboratory's CIT cluster, which is supported by
National Science Foundation Grants PHY-0757058 and PHY0823459. ROS
acknowledges support from NSF Grant No. AST-1909534, NSF Grant No.
PHY-2012057, and the Simons Foundation. MQ acknowledges the
computational resources provided by Habib University.

\appendix

\section{Synthetic catalog and recovery procedure}
\label{ap:mock_workflow}

For the spinning, eccentric BBH and multisource tests, the GWKokab
generator draws a Poisson count with mean $\mu(\pvec_*)$ at the
injected population. It draws a larger intrinsic pool and resamples
with weights proportional to $\mathrm{VT}(\svec)$, approximating the
distribution of detected sources,
$\prob(\svec\mid{\rm det},\pvec_*)\propto
\prob(\svec\mid\pvec_*)\mathrm{VT}(\svec)$. Distance and
orientation are integrated into the sensitivity model. The BBH
sensitivity depends on masses and aligned spins and is taken to be
independent of eccentricity. The full-model test draws 10,074
intrinsic sources and retains 74 detections, all analyzed by both
methods. The multisource test reuses the selected catalog of
Ref.~\cite{krnm-3vrf}; the two toy catalogs contain 115 and 104 events.

Selection in the full-model and multisource recovery uses a neural
interpolant of a semianalytical $\mathrm{VT}$ model calibrated with
O3 sensitivity injections. At each trial population we average
$\mathrm{VT}$ over 2000 model draws per component and multiply
by its rate to estimate the expected count. These calibration
injections and integration draws are separate from the mock
catalog. The discrete and continuous analyses use the same catalog,
population model, and expected-count calculation; they differ in the
numerical evaluation of the event integrals. The real-event catalog
comparisons in Secs.~\ref{sec:real_event_recovery} and
\ref{sec:gwtc5_recovery} instead use their catalog-specific selection
injections; the synthetic O3 interpolant is not used for those results.

\section{Finite boundaries: Truncated versus diffusive error models}
\label{ap:boundaries}
The mass--spin--eccentricity test in Sec.~\ref{sec:full_model} uses
the truncated spin and eccentricity errors and discarded unphysical
mass draws specified in Sec.~\ref{sec:sub:syn_population}.
The reflecting diffusion kernel below provides a different error
model, with zero flux at the boundaries. We use it consistently for
data generation and likelihood evaluation in the worked example
below; it does not replace the error prescription in
Sec.~\ref{sec:full_model}.

In the main text, we generate synthetic data for ${\cal M}_c$, $\eta$,
and the component spins by (a) drawing values from a population,
(b) adding perturbations drawn from a prescribed normal measurement-
error distribution, and (c) using the known error distribution to
generate posterior samples. For code testing and validation, we
occasionally replace step (c) with the exact likelihood. This
procedure requires an explicit treatment of boundary conditions. For
example, $\eta<1/4$, so random draws that
produce $\eta>1/4$ must be either rejected (producing a truncated
normal) or reflected (to mimic zero-flux boundary
conditions while changing the shape of the assumed error model).
Thus, the choice of boundary condition changes the error model and
can affect downstream conclusions near a boundary.

In this Appendix, we describe the posterior-generation process,
introduce a diffusive error model that is particularly well suited to
finite boundaries and large measurement errors, and discuss its
implications. For simplicity, we restrict the discussion to one
dimension,
$x\in [-L/2,L/2]$ (e.g., similar to $\chi_{i,z}$)
although the same reasoning applies to $\eta$, eccentricity, and
other bounded parameters. We do not assume periodicity.

Given a generative population model $p(x)$, an error model
$p_e(y\mid x)$ produces a data distribution $p_d$ for $y$:
\begin{align}
	p_d(y) & = \int p_e(y|x)p(x) dx
\end{align}
For a fixed value of $y$, we use $p_e$ to construct a fiducial
posterior relative to a fiducial prior $p_{\rm prior}(x)$:
\begin{align}
	p_p(z) & = \frac{ p_e(y|z)p_{\rm prior}(z) }{\int dz p(y|z)p_{\rm prior}(z)}
\end{align}
The fiducial posterior $p_p(z)$ is usually represented by a finite
number of independent samples drawn from the distribution above. If
$p_{\rm prior}$ is uniform and $p_e$ is normal, then
$p_e(y\mid z)$ is symmetric in $y$ and $z$. We can therefore draw
efficiently from the normal distribution $p_e$ to generate posterior
samples, given the single draw from $p_e$ used to select $y$.
In the absence of boundaries, $p_e$ can be interpreted as generating
a small offset, $y=x+\Delta x$, where the distribution of
$\Delta x$ is centered at zero and has a specified standard
deviation.

For models with finite boundaries, however, $p_e$ cannot be exactly
normal; it may instead be a truncated normal distribution. The
closed-form expression for a truncated normal makes clear that the
normalization depends on $\mu$. Consequently, we cannot assume that
$p_e(y\mid z)$ is symmetric in $y$ and $z$. In other words,
$p_e(y\mid x)$, considered as a function of $y$, is a truncated
normal distribution, whereas $p_e(y\mid z)$, considered as a function
of $z$, is not.

\subsection{Approach 1: Use a truncated normal error model}
One option is to use a truncated normal distribution as the error
model. This distribution is typically characterized by $\mu=0$ and
a nominal width $\tilde{\sigma}$, provided that we use
$p_e(y\mid z)$ correctly to draw $z$ for the synthetic posterior.
Specifically, we adopt $p_e(y\mid x)=p_t(y\mid x,\sigma)$, where
$p_t(x\mid\mu,\sigma)$ denotes a truncated normal distribution on
$[-1,1]$:
\begin{align}
	p_t(x|\mu,\sigma) & =
	e^{-(\mu-x)^2/2\sigma^2} \frac{1}{Z(\mu,\sigma)\sqrt{2\pi \sigma^2}}
	\\
	Z(\mu,\sigma)     & = \Phi\left( \frac{1-\mu}{\sigma} \right)  -
	\Phi\left( \frac{-1-\mu}{\sigma} \right)
\end{align}
Far from the boundaries, this procedure reduces to the conventional
normal error model.

Although this approach provides a closed-form expression for
$p_e(y\mid x)$, it is inconvenient for both sampling and analysis.
As a function of $x$, the expression is not normalized and therefore
requires costly, $y$-dependent calculations to generate fiducial
posterior samples efficiently. For analytic calculations, the
normalization can make the resulting expressions intractable and
breaks the useful symmetry between $x$ and $y$ that is present in the
diffusive approach below. Even forward calculations are difficult:
when the population model $p(x)$ is a truncated normal distribution,
we cannot obtain a closed-form expression for the data distribution
$p_d(y)$.

\subsection{Approach 2: Use a diffusive error model}
Alternatively, we can apply the error diffusively through a random
walk with a characteristic diffusion time. Specifically, we assume
that $p_e(y\mid x)=p(y\mid x,\tau)$ satisfies a diffusion equation
with $\delta$-function initial data and zero-flux boundary conditions:
\begin{align}
	\partial_\tau p = D \nabla^2 p
\end{align}
such that $2D \tau =\sigma_{\rm net}^2$.  This approach consistently
applies a well-defined amount of jitter to every
point in the parameter space, including points near boundaries.
This Markovian approach implies that the data distribution $p_d$ also
evolves according to the same diffusion process.

This approach is highly efficient for data generation: we apply
random walks with reflecting boundary conditions. For example, a
synthetic parameter $y$ is obtained by applying a random walk to a
draw of $x$. Because the symmetry condition holds, posterior samples
of $z$ are generated similarly by diffusing from $y$.

Although data generation is straightforward, the method produces
less familiar expressions for $p_e(y\mid x)$ in Bayes' theorem. On an
infinite interval, the diffusion kernel is a Gaussian with standard
deviation $\sqrt{2D\tau}$. If this kernel can be convolved with
$p(x)$, the data distribution may be analytically tractable. On a
half-infinite interval with a single relevant boundary, such as
$x_-=-L/2$, diffusion can be implemented by reflection. For example,
the standard infinite-domain kernel can be applied to the extended
distribution $p(2x_- - x)+p(x)$. On a bounded domain where both limits
are important, the diffusion kernel $p(y\mid x,\tau)$ can be
calculated using series methods. It can be expressed either as an
infinite series of images reflected across both boundaries or as a
Fourier series. In either representation, only a small number of
terms is generally needed to estimate $p(y\mid x,\tau)$, and hence
$p_d(y)$, efficiently.
\begin{widetext}
	On the interval $x,y\in[x_-,x_+]$ of length $L=x_+-x_-$, the two
	series solutions for the diffusive error model can be written as
	\begin{align}
		p_e(y|x,\tau) & =  \frac{1}{\sqrt{2\pi \sigma_{\rm net}^2}}
		\sum_{n=-\infty}^{\infty}\left[
			e^{-(y-2Ln-2x_-+x)^2/2\sigma_{\rm net}^2} +     e^{-(y-2Ln -
					x)^2/2\sigma_{\rm net}^2}
			\right]
		\\
		              & = \frac{1}{L} + \frac{2}{L} \sum_{n=1}^\infty \cos
		\frac{n\pi}{L}(y-x_-) \cos \frac{n\pi}{L}(x-x_-)
		e^{-(n\pi/L)^2\sigma_{\rm net}^2/2}
	\end{align}
	where $\sigma_{\rm net}^2=2 D\tau$.
	As noted above, both expressions are symmetric under the interchange
	$y\leftrightarrow x$, as required by the diffusion model. The first
	series is particularly efficient when $\sigma_{\rm net}$ is small or
	when $x$ and $y$ are close to a boundary. The second series is
	particularly efficient when $\sigma_{\rm net}$ is large. In either
	case, only a few terms are usually required for an accurate estimate.
	In principle, the first series could be summed using Jacobi
	$\theta$ functions with suitable arguments. In practice, however,
	the lack of numerical or autodifferentiable implementations of these
	functions limits the utility of such a resummation.
\end{widetext}
For a population on the bounded interval, the cosine moments
\begin{align}
 c_n(\tau)&=\int_{x_-}^{x_+}p(x,\tau)
 \cos\!\left[\frac{n\pi(x-x_-)}{L}\right]\,\mathrm{d}x
\end{align}
evolve as $c_n(\tau)=c_n(0)\exp[-D\tau(n\pi/L)^2]$.
The uniform mode $c_0=1$ is preserved. This representation propagates
the population through the same reflecting kernel used for individual
events. The free-space moment-generating formula
$M_{p_d}(K)=\exp(D\tau K^2)M_p(K)$ does not apply on a bounded
reflecting interval, where integration by parts introduces boundary
terms.

\subsection{Worked reflecting-diffusion example}
\label{ap:diffusion_example}

Write $K_s$ for the reflecting kernel above on $[0,1]$, with
diffusion scale $s$. We generate 256 sources from
$p(x\mid s)=K_s(x\mid\mu)$, fixing $\mu=0.02$ and injecting
$s=0.02$. For each source we draw one datum
$y\sim K_{\sigma_e}(y\mid x)$ with $\sigma_e=0.06$. Both stages
can be sampled by folding a Gaussian draw back into $[0,1]$.
The center $\mu$ and scale $s$ specify the heat kernel; they are
not the mean and standard deviation after reflection.
Selection is unity and the catalog size is fixed.

\begin{figure*}[t]
\centering
\includegraphics[width=0.98\textwidth]{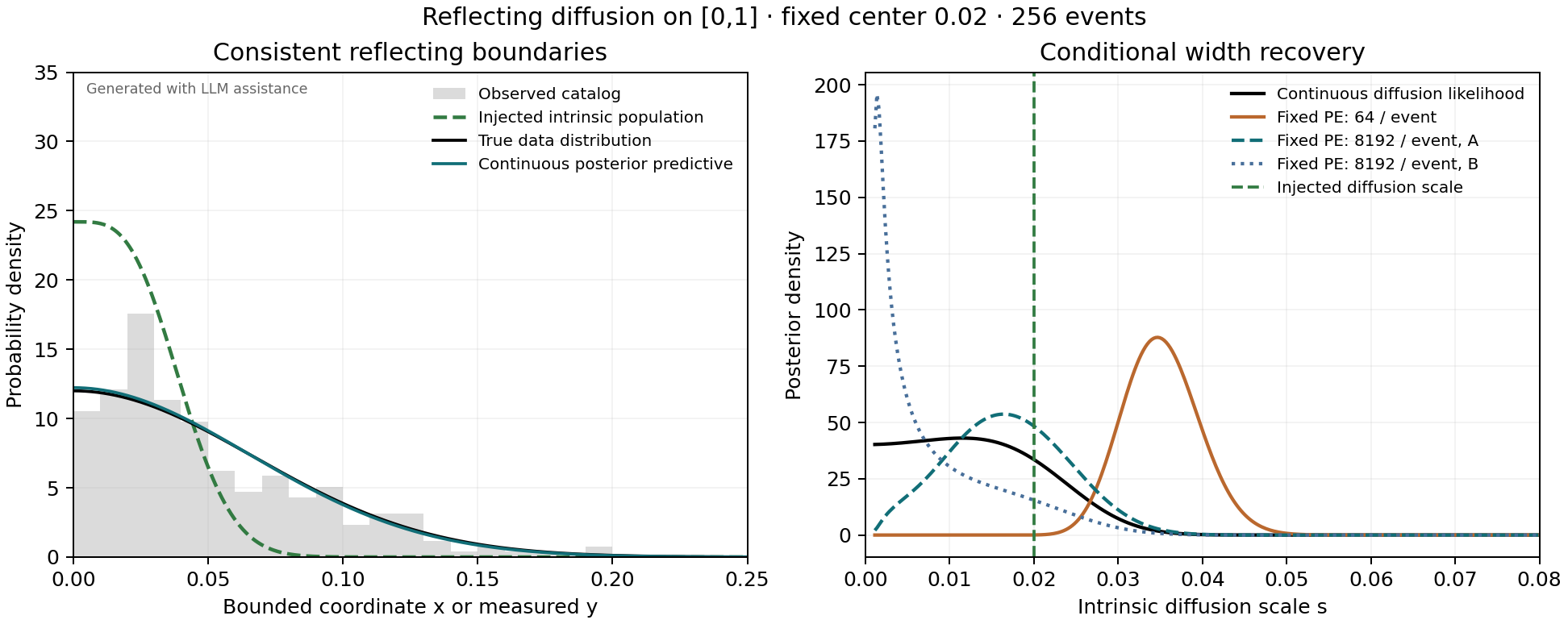}
\caption{Worked example using reflecting diffusion for the intrinsic
population, measurement errors, and event posterior samples.
Left: the injected intrinsic population, noisy catalog, exact data
distribution, and continuous posterior predictive distribution of
a future measured datum $y$.
Right: recovery of the intrinsic diffusion scale $s$ using the
exact continuous event integral or fixed PE clouds. The dashed
vertical line marks the injected scale $s=0.02$; the kernel center
is fixed at $\mu=0.02$ and the measurement scale is $0.06$.
The two 8192-sample PE banks still differ from the continuous
reference and from each other. These are diffusion scales, not the reflected
distributions' moment standard deviations. \llmfig}
\label{fig:diffusion_toy_recovery}
\end{figure*}

With a uniform reference prior on $[0,1]$, symmetry and
normalization give $p(x\mid y)=K_{\sigma_e}(x\mid y)$, so PE
samples are generated by the same reflection procedure, conditional
on the observed datum rather than on the latent source.
The semigroup property gives an independent exact event integral,
\begin{align}
 \int_0^1 K_{\sigma_e}(y\mid x)K_s(x\mid\mu)\,\mathrm{d}x
 &=K_{\sqrt{s^2+\sigma_e^2}}(y\mid\mu).
\end{align}
We infer only $s$, using a uniform prior on $[0.001,0.08]$.
Figure~\ref{fig:diffusion_toy_recovery} compares the resulting
continuous likelihood with finite-cloud averages using 64 PE
samples per event and two independent 8192-sample banks.
This gives an independently checkable benchmark of the complete
inference procedure near a physical boundary. The symmetry of the
kernel makes event-posterior sampling straightforward, and its
semigroup property supplies an exact answer for the population
integrals. The population and error models are therefore fully
specified, so discrepancies between the recovery methods isolate
finite-sample integration error. The exact 90\% credible interval
contains the injected scale. With 64 posterior samples per event,
the interval excludes it and favors a larger width. Both
8192-sample estimates include the injection, but their posteriors
still differ from each other and from the exact result. The value of
the diffusion model is that it preserves the boundary and
normalization at every stage while allowing accurate continuous
integration, without relying on a fixed posterior cloud. It is a
useful test of boundary-aware inference, not a claim of a resolved
nonzero population width.

\section{Analytic-fit eligibility and low-mass coverage}
\label{ap:fit_coverage}

The event-selection cut and the choice of likelihood representation
are separate. The original 153-event cohort follows the RDL
false-alarm-rate cut of $1\,\mathrm{yr}^{-1}$; the extension adds the
104 O4b BBHs in the GWTC-5 population release. We impose no mass cut
to choose which events receive analytic likelihoods. The
$5\,M_\odot$ secondary-mass floor instead defines the integration
domain of this conditional population comparison.

Our primary comparison uses the exact adopted RDL fits for all 153
original events and 97 matching four-dimensional long-EM fits for the
extension. The original adopted archive and the public release are
distinct: the latter includes revised fits and reference analyses,
so membership in a particular public collection is not an eligibility
requirement. Model and reference hashes identify the adopted inputs.
Every included fit passes the native-posterior shape criteria:
marginal Jensen--Shannon divergence at most 0.03 bits and axis
Kolmogorov--Smirnov distance at most 0.1. To probe correlations,
we require Jensen--Shannon divergence at most 0.04 bits in each of
eight one-dimensional projections along fixed directions in the
standardized $(\mathcal M_{c,\rm src},q,\chi_{\rm eff},z)$ space.
We also require divergence at most 0.06 bits for the mass-ratio
distribution in each of three equal-probability redshift intervals
of the reference posterior.
For the extension we require these checks on three validation
splits. These criteria assess reproduction of the reference posterior;
they do not certify arbitrary population reweightings.

For the holdout test, roughly one fifth of the reference intrinsic
draws are reserved for validation and the fit is trained on the rest
\cite{hemrattaphanLikelihoodReconstruction}. All rows from one
intrinsic draw stay on the same side of the split. When these group
identifiers are unavailable, we split by row, so training and
validation rows can share intrinsic parameters. Each validation
split entails a separate refit of the same event; it tests sensitivity
to the training and validation sample choice.

The restricted arm in Figs.~\ref{fig:gwtc4_mass_recovery}
and~\ref{fig:gwtc5_mass_recovery} requires an exact match to one of
two public base8 collections for the original events. It applies
the same shape criteria to the native posterior and six population
reweightings. It also requires reference effective sample size at
least 80, proposal effective sample size at least 100, absolute population
log-response error at most 0.2 nats, and absolute split-half log-response
difference at most 0.2 nats. Here the response is the change in the
log event integral relative to the first trial population. Its error
compares the fit with held-out PE; its split-half difference compares
responses from the two halves of the fit's integration draws.
For each weighting, the redshift intervals used for the conditional
mass-ratio check are recomputed from the reweighted reference.
For the original collections we use the published validation
split; for the extension we require all three splits. These criteria select 67
original fits, including GW150914, and 67 extension fits.
These diagnostic thresholds do not bound the population posterior
error. The paired comparison tests the combined effect of the
diagnostic criteria and public-release matching on the same catalog.
Of the 153 original fits, 105 satisfy the full set of diagnostics.
The restricted analysis uses the 67 that also match a public
collection. The other diagnostic outcomes remain recorded for the
enlarged set.

\begin{figure*}[t]
\centering
\includegraphics[width=0.98\textwidth]{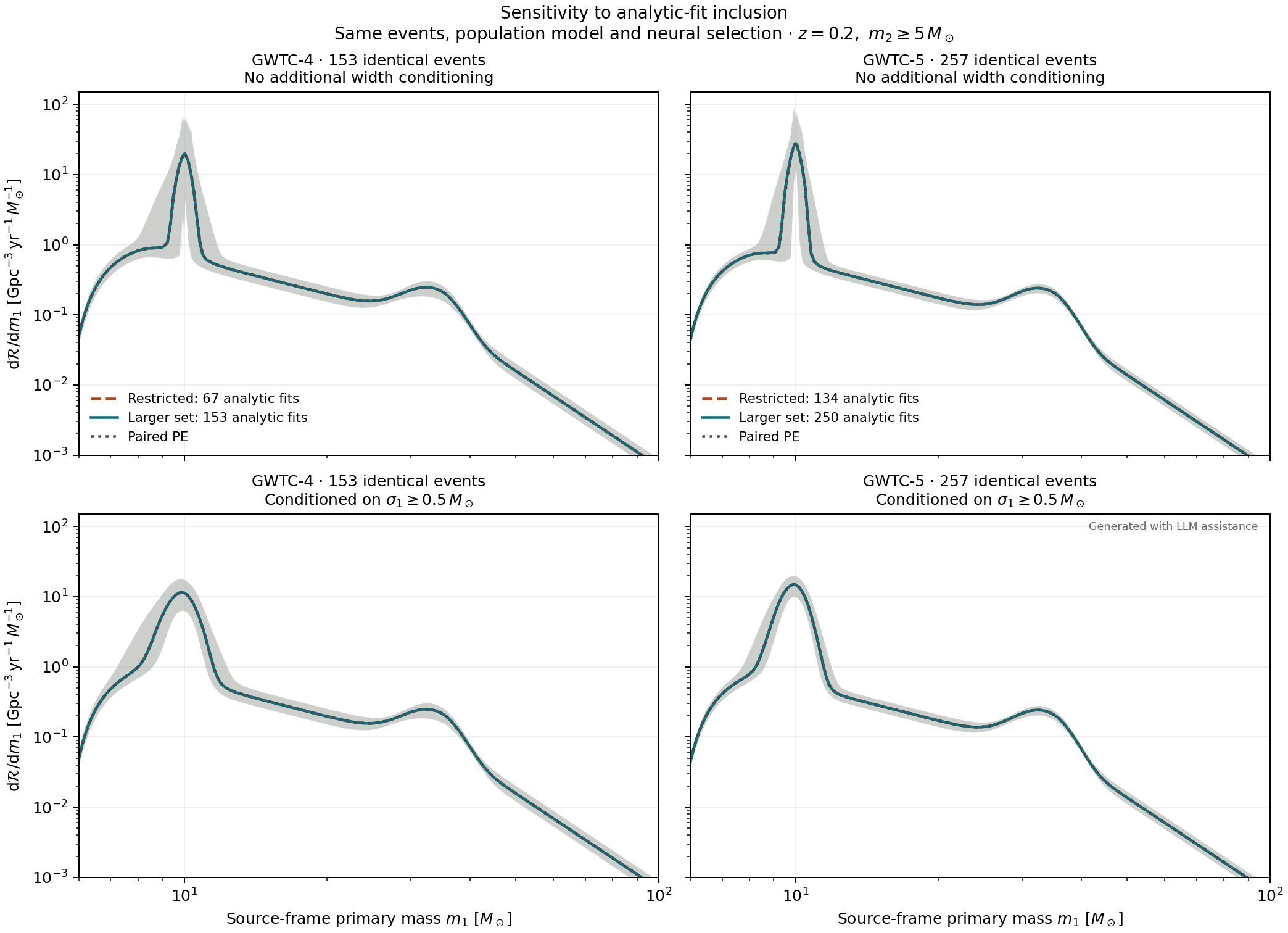}
\caption{Sensitivity to analytic-fit inclusion with neural selection.
The GWTC-4 curves use 0, 67, and 153 fits (left). The GWTC-5 curves
use 0, 134, and 250 fits (right). Any event without a fit in a given
arm uses its matched PE reference.
Within each catalog, all curves use the same neural selection model and fixed nuisance
parameters. Upper panels include all tested widths; lower panels
condition on $\sigma_1\geq0.5\,M_\odot$. Bands show pointwise 90\%
credible intervals from joint shape and rate-amplitude draws.
The restricted and enlarged results nearly coincide, while the
narrow-width limitations in the text remain. \llmfig}
\label{fig:fit_inclusion_recovery}
\end{figure*}

Figure~\ref{fig:fit_inclusion_recovery} repeats the three-arm
comparison on the identical event catalogs and numerical grid using
a common neural selection model.
Increasing the analytic-fit count from 67 to 153 changes the median
rate at $10\,M_\odot$ from 19.53 to 19.51; increasing it from 134 to
250 changes that rate from 27.65 to 27.70, in
$\mathrm{Gpc}^{-3}\,\mathrm{yr}^{-1}\,M_\odot^{-1}$.
These full-bank results include all tested widths. The fit-inclusion
choice has little observed effect in this conditional test.
This agreement does not resolve the narrowest
widths or validate the added fits for other population models.

Table~\ref{tab:fit_mass_coverage} summarizes coverage using each
event's weighted reference-posterior median source-frame primary
mass. These bins describe coverage and are not selection criteria.
In the enlarged set, all 34 original events with median
$m_1<20\,M_\odot$ have analytic fits. In the extended cohort,
62 of 64 such events have fits.
The two low-mass PE fallbacks are GW240520\_213616 and
GW240627\_131622, with median primary masses 11.3 and
$12.1\,M_\odot$. Their posterior probabilities of $m_1<15\,M_\odot$
are 97\% and 85\%, respectively, so they do overlap the first-peak
region. The other five fallbacks are GW240420\_175625,
GW240618\_071627, GW240621\_195059, GW240621\_214041, and
GW240921\_201835. These seven lack a matching four-dimensional
long-EM fit in the adopted archive; the primary comparisons retain
them with PE likelihood integrals.

As a completion check, we constructed matching RDL fits for these
seven reference analyses. All pass the native shape criteria on
three splits, permitting an additional calculation with analytic
fits for all 257 events. Holding the event catalog and injection
selection fixed, this changes the median rate at $10\,M_\odot$
from 45.17 to 44.32 when all tested widths are included, and from
14.66 to 14.67 when conditioning on
$\sigma_1\geq0.5\,M_\odot$. Replacing all seven PE representations
together therefore has little effect in this test.

\begin{table}[t]
\centering
\caption{Analytic-fit coverage by weighted reference-posterior median
primary mass, in $M_\odot$. The restricted and enlarged columns count
analytic fits in the corresponding analyses of
Figs.~\ref{fig:gwtc4_mass_recovery} and~\ref{fig:gwtc5_mass_recovery},
using the criteria in this appendix. Events without an analytic fit
in a given analysis use their matched PE reference.}
\label{tab:fit_mass_coverage}
\begin{tabular}{lrrr}
\hline
Median $m_1$ & All events & Restricted & Enlarged\\
\hline
\multicolumn{4}{c}{Original cohort}\\
$<10$ & 4 & 1 & 4\\
$10$--$15$ & 24 & 15 & 24\\
$15$--$20$ & 6 & 2 & 6\\
$\geq20$ & 119 & 49 & 119\\
\hline
\multicolumn{4}{c}{Extended cohort}\\
$<10$ & 8 & 5 & 8\\
$10$--$15$ & 45 & 34 & 43\\
$15$--$20$ & 11 & 6 & 11\\
$\geq20$ & 193 & 89 & 188\\
\hline
\end{tabular}
\end{table}

\section{Detection volumes}
\label{ap:VT}
For populations in which $\comp$ is independent of redshift, source
orientation, and sky location, the integral defining $\mu$ can be
rearranged to marginalize over these degrees of freedom:
\begin{equation}\label{eq:expected_rates}
	\mu (\pvec) = T {\cal R}(\svec) \int V(\svec) \prob(\svec|\pvec)
	\sqrt{g_\svec}  d\svec,
\end{equation}
where $V(\svec)$ is a precomputed effective volume that includes any
necessary redshift factors, such as $1/(1+z)$ for converting
source-frame time to detector-frame time.

For example, a simple estimate of $V(\svec)$ can be obtained by
adopting a fixed signal-to-noise-ratio threshold in a single
detector. For a nonprecessing model dominated by a single
$\ell=|m|=2$ mode, the orientation-averaged sensitive three-volume is
\begin{equation}\label{eq:V}
	V(\svec) =
	\int
	P(> D(z)/D_h(\svec))
	\frac{\mathrm{d}V_{\mathrm{c}}}{\mathrm{d}z}
	\frac{\mathrm{d}z}{1+z},
\end{equation}
where $D(z)$ is the luminosity distance for redshift $z$;
$D_h(m_1(1+z),m_2(1+z))$ is the horizon distance to which the source
can be seen; $V_{\mathrm{c}}$ is the comoving volume;
$z$ is the redshift of the merger event; and the cumulative
distribution $P(>w)$ is the survival probability of the
orientation projection factor $w=\rho/\rho_{\rm opt}$, with its
isotropic sky, polarization, and inclination measure normalized to
unity. The threshold is $D(z)/D_h$.
Although this estimate and its generalizations can be computed
analytically, more sophisticated estimates are generally derived by
applying a real search pipeline to a fixed but large number of
candidate signals and recording which signals are recovered.


\paragraph{Catalog-calibrated neural selection.}
For Sec.~\ref{sec:continuous_selection}, let $H_{\rm cat}(m_1)$
be the effective volume obtained by integrating the fixed nuisance
densities and the secondary-mass cut in the catalog-specific
injection weights. The selection integral is then
$A(\Lambda)=\int p_1(m_1\mid\Lambda)S_1(m_1)
H_{\rm cat}(m_1)\,\mathrm{d}m_1$, where $p_1$ is the pre-taper
primary-mass mixture and $S_1$ is its taper. Observation time is
applied once when converting this volume to an expected count.
We fit $\ln H_{\rm cat}$ in 180 logarithmic mass bins using a
GWKokab neural network with two 32-unit ReLU hidden layers and a
fixed 1500-epoch AdamW schedule. The log-density loss is weighted
by effective injection count, capped at 500; bins require an
effective count of 15.
The analytic primary taper is restored after fitting. Independently
trained injection halves are compared with unbinned held-out
integrals for 25 Gaussian populations. The largest discrepancies
are about three estimated standard errors of the held-out injection
integrals; these errors do not include uncertainty in the neural fit.
Each component integral uses
two independent scrambled Sobol sets of 65,536 population draws,
with inverse-CDF sampling for the truncated Gaussians. Across the
full tested grid, the two integration replicas differ by less than
0.006 in $N\ln A$ for a fixed fitted curve. This checks numerical
integration of the curve. The fitted halves share architecture and binning,
so their agreement does not exclude a common interpolation bias.
Selection-calibration uncertainty is checked through these comparisons,
rather than marginalized in the reported posteriors.

\subsection{Neural and injection-based selection in recovery}
\label{ap:selection_recovery}

\begin{figure*}[t]
\centering
\includegraphics[width=0.98\textwidth]{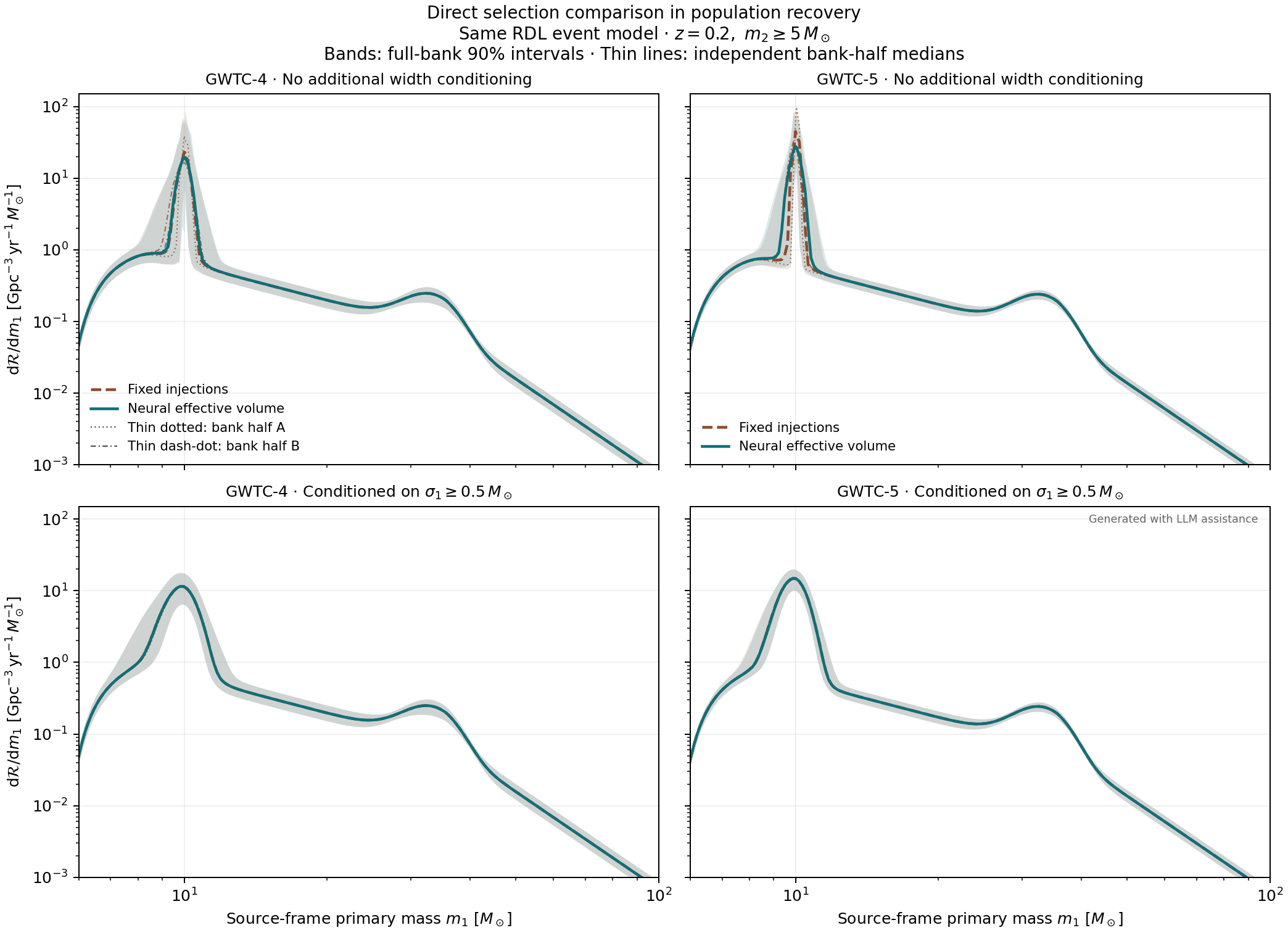}
\caption{Direct comparison of recovery using neural selection and
fixed-injection selection for GWTC-4 (left) and GWTC-5 (right).
The event likelihoods and population model are held fixed. Thick
curves and bands show full-bank medians and pointwise 90\%
credible intervals for the joint shape and rate posterior; thin
dotted and dash-dotted curves show the two bank-half medians.
Upper panels apply no additional width conditioning and lower
panels condition both methods on $\sigma_1\geq0.5\,M_\odot$.
The local width prior is $0.05$--$10\,M_\odot$. The nuisance
parameters remain fixed, and neural calibration uncertainty is
not marginalized. The narrow-width and interpolation limitations
discussed in Sec.~\ref{sec:continuous_selection} still apply. \llmfig}
\label{fig:neural_injection_recovery}
\end{figure*}

Figure~\ref{fig:neural_injection_recovery} directly compares the
recovered rate density when only the selection estimator changes.
Both calculations use the same RDL/PE event representation,
catalog, population model, integration grid, and joint shape and
rate-amplitude sampling. The upper panels include all tested widths;
the lower panels condition both calculations on
$\sigma_1\geq0.5\,M_\odot$. Thin curves show the medians obtained
from disjoint event-integration draws and selection-injection sets,
with separately
trained neural fits for the latter. They use all catalog events
and are not separate half-catalog analyses.

\clearpage
\section{Use of large language models}
\label{app:llm}

R.~O'Shaughnessy was the only author who used large language models
(LLMs) for this work. He directed OpenAI Codex using GPT-6.1 Sol
(model identifier \texttt{gpt-6.1-sol}) to implement and run the
real-event recovery calculation and selection check, generate
Figs.~\ref{fig:good_event_fit}, \ref{fig:gwtc4_mass_recovery},
\ref{fig:gwtc5_mass_recovery}, \ref{fig:diffusion_toy_recovery},
\ref{fig:neural_injection_recovery}, and \ref{fig:fit_inclusion_recovery}
from the numerical output, and
draft text reporting the results. He chose the comparison and
validation criteria and reviewed the figure and results. He also
used GPT-6.2 Astra (model identifier \texttt{gpt-6.2-astra}) for an
adversarial scientific and clarity review of the appendix additions. The
checks included paired PE analyses, disjoint integration banks,
and grid refinement. The workflow and its documentation and review
procedures are described in
Ref.~\cite{whitepaper-transparent-agentic-science}. The authors take
responsibility for the content of the paper.

\bibstyle{apsrev4-2}
\makeatletter
\let\pre@bibdata\@empty
\makeatother
\bibliography{refs}

\end{document}